\documentclass[sn-mathphys,Numbered]{sn-jnl}

\usepackage{graphicx}%
\usepackage{multirow}%
\usepackage{amsmath,amssymb,amsfonts}%
\usepackage{amsthm}%
\usepackage{mathrsfs}%
\usepackage[title]{appendix}%
\usepackage{textcomp}%
\usepackage{manyfoot}%
\usepackage{booktabs}%
\usepackage{algorithm}%
\usepackage{algorithmicx}%
\usepackage{algpseudocode}%
\usepackage{listings}%

\usepackage{bm}%
\usepackage{adjustbox}%
\usepackage[table]{xcolor}%
\usepackage{siunitx}%
\DeclareSIUnit\angstrom{\text {Å}}%
\usepackage{subcaption}%

\usepackage{geometry}
\newcommand{\one}[1]{#1}%
\newcommand{\two}[1]{#1}%
\newcommand{\three}[1]{#1}%
\newcommand{\best}[1]{\bm{#1}}%

\begin{document}

\title[Article Title]{Structure-based Drug Design with Equivariant Diffusion Models}

\author*[1]{\fnm{Arne} \sur{Schneuing}}\email{arne.schneuing@epfl.ch}
\equalcont{These authors contributed equally to this work.}

\author[2]{\fnm{Charles} \sur{Harris}} 
\equalcont{These authors contributed equally to this work.}

\author[3]{\fnm{Yuanqi} \sur{Du}} 
\equalcont{These authors contributed equally to this work.}

\author[2]{\fnm{Kieran} \sur{Didi}} 
\author[2,9]{\fnm{Arian} \sur{Jamasb}} 
\author[1]{\fnm{Ilia} \sur{Igashov}} 
\author[4]{\fnm{Weitao} \sur{Du}} 
\author[3]{\fnm{Carla} \sur{Gomes}} 
\author[2,10]{\fnm{Tom L.} \sur{Blundell}} 
\author[2,5]{\fnm{Pietro} \sur{Lio}} 
\author[6,11]{\fnm{Max} \sur{Welling}} 
\author[7,8]{\fnm{Michael} \sur{Bronstein}} 
\author*[1]{\fnm{Bruno} \sur{Correia}}\email{bruno.correia@epfl.ch}

\affil[1]{\orgname{École Polytechnique Fédérale de Lausanne}, \orgaddress{\city{Lausanne}, \country{Switzerland}}}
\affil[2]{\orgname{University of Cambridge}, \orgaddress{\city{Cambridge}, \country{UK}}}
\affil[3]{\orgname{Cornell University}, \orgaddress{\city{Ithaca}, \country{USA}}}
\affil[4]{\orgname{Chinese Academy of Mathematics and System Science}, \orgaddress{\city{Beijing}, \country{China}}}

\affil[5]{\orgname{University of Rome ``La Sapienza''}, \orgaddress{\city{Rome}, \country{Italy}}}
\affil[6]{\orgname{Microsoft Research AI4Science}, \orgaddress{\city{Amsterdam}, \country{Netherlands}}}
\affil[7]{\orgname{University of Oxford}, \orgaddress{\city{Oxford}, \country{UK}}}
\affil[8]{\orgname{AITHYRA Institute}, \orgaddress{\city{Vienna}, \country{Austria}}}

\affil[9]{\orgname{Current affiliation: Prescient Design, Genentech}, \orgaddress{\city{Basel}, \country{Switzerland}}}
\affil[10]{\orgname{Current affiliation: Heart and Lung Research Institute,
University of Cambridge}, \orgaddress{\city{Cambridge}, \country{UK}}}
\affil[11]{\orgname{Current affiliation: University of Amsterdam}, \orgaddress{\city{Amsterdam}, \country{Netherlands}}}

\abstract{
Structure-based drug design (SBDD) aims to design small-molecule ligands that bind with high affinity and specificity to pre-determined protein targets. 
%
Generative SBDD methods leverage structural data of drugs in complex with their protein targets to propose new drug candidates.
These approaches typically place one atom at a time in an autoregressive fashion using the binding pocket as well as previously added ligand atoms as context in each step. Recently a surge of diffusion generative models has entered this domain which hold promise to capture the statistical properties of natural ligands more faithfully.
%
However, most existing methods focus exclusively on bottom-up \textit{de novo} design of compounds or tackle other drug development challenges with task-specific models.
The latter requires curation of suitable datasets, careful engineering of the models and retraining from scratch for each task.
%
Here we show how a single pre-trained diffusion model can be applied to a broader range of problems, such as off-the-shelf property optimization, explicit negative design, and partial molecular design with inpainting.
%
We formulate SBDD as a 3D-conditional generation problem and present DiffSBDD, an SE(3)-equivariant diffusion model that generates novel ligands conditioned on protein pockets.
Our \textit{in silico} experiments demonstrate that DiffSBDD captures the statistics of the ground truth data effectively. 
Furthermore, we show how additional constraints can be used to improve the generated drug candidates according to a variety of computational metrics. 
%
These results support the assumption that diffusion models represent the complex distribution of structural data more accurately than previous methods, and are able to incorporate additional design objectives and constraints changing nothing but the sampling strategy.
%
We anticipate that our findings may contribute to accelerate progress on several computational drug design frontiers as more powerful distribution learners emerge, that can be inserted into our flexible framework.
}


\maketitle

\setlength{\parskip}{\baselineskip}%
\setlength{\parindent}{0pt}%

\section{Introduction}\label{sec:intro}

The rational design of small-molecules with drug-like properties remains an outstanding challenge in both fundamental and biopharmaceutical research.
Structure-based drug design (SBDD) aims to find  small-molecule ligands that bind to specific three-dimensional sites in proteins with high affinity and specificity \cite{anderson2003SBDD3}. Traditionally, SBDD campaigns are usually initiated either by high-throughput experimental or virtual screening \cite{lyne2002structure, shoichet2004virtual} of large chemical databases. Generally, these approaches are expensive and time-consuming, but they also  restrict the exploration of the chemical space to  previously studied molecules, with a further emphasis usually placed on commercial availability~\cite{irwin2005zinc}. Moreover, the optimization of initial lead molecules is often a biased process, with significant reliance on human intuition \cite{ferreira2015SBDD2}.
Recent advances in geometric deep learning, especially in modelling geometric structures of biomolecules~\cite{bronstein2021geometric,atz2021geometric,khakzad2023new}, provide a promising direction for SBDD~\cite{Gaudelet2021}. Despite remarkable progress in the use of deep learning as surrogate docking models~\cite{lu2022tankbind, stark2022equibind, corso2022diffdock}, deep learning-based design of ligands that bind to target proteins remains an overarching problem in molecular modeling. Early attempts have been made to represent molecules as atomic density maps, with variational auto-encoders generating new atomic density maps corresponding to novel molecules~\cite{ragoza2022generating}. However, it is nontrivial to map atomic density maps back to molecular space, requiring an additional atom-fitting stage. An alternative is to represent molecules as 3D graphs with atomic coordinates and types which naturally circumvents the post-processing steps. \citet{li2021structure} proposed an autoregressive generative model to sample ligands given the protein pocket as a conditioning constraint. \citet{peng2022pocket2mol} improved this method by using an $E(3)$-equivariant graph neural network which respects rotation and translation symmetries in 3D space.
Similarly, \citet{drotar2021structure} and \citet{liu2022generating} used autoregressive models to generate atoms sequentially and incorporate angles during the generation process. 
However, the main premise of sequential generation methods may not hold in real scenarios, since it imposes an artificial ordering scheme in the generation process and, as a result, the global context of the generated ligands may be lost. 
Very recently, a number of diffusion models have been put forward for target-specific molecule design~\citep{guan20233d, lin2022diffbp, guan2023decompdiff, xu2023geometric, weiss2023guided}. These models place all atoms simultaneously, allowing them to reason about the whole molecule at once and typically enabling faster sampling. While this class of models has already shown great promise in \textit{de novo} ligand generation, their potential in other parts of the drug design pipeline has not been thoroughly explored.

In this study, we propose DiffSBDD, an $SE(3)$-equivariant 3D-conditional diffusion model for SBDD that respects translation, rotation, and permutation symmetries. 
With a free \textit{de novo} generation benchmark, we first establish that the diffusion model captures molecular properties of real molecules more accurately than previously popular autoregressive models.
Secondly, we show how the design space can be effectively constrained based on prior knowledge to improve the sample quality. Our inpainting-inspired approach furthermore allows us to tackle a diverse set of molecular design problems, such as scaffold hopping/elaboration and fragment growing/merging, all without the need to retrain new models.
Lastly, we demonstrate how pre-trained DiffSBDD models can be used out-of-the-box to optimize arbitrary molecular properties via a noise/denoise scheme in combination with an evolutionary algorithm.

\begin{figure*}[p!]
    \centering
    \includegraphics[width=0.95\textwidth]{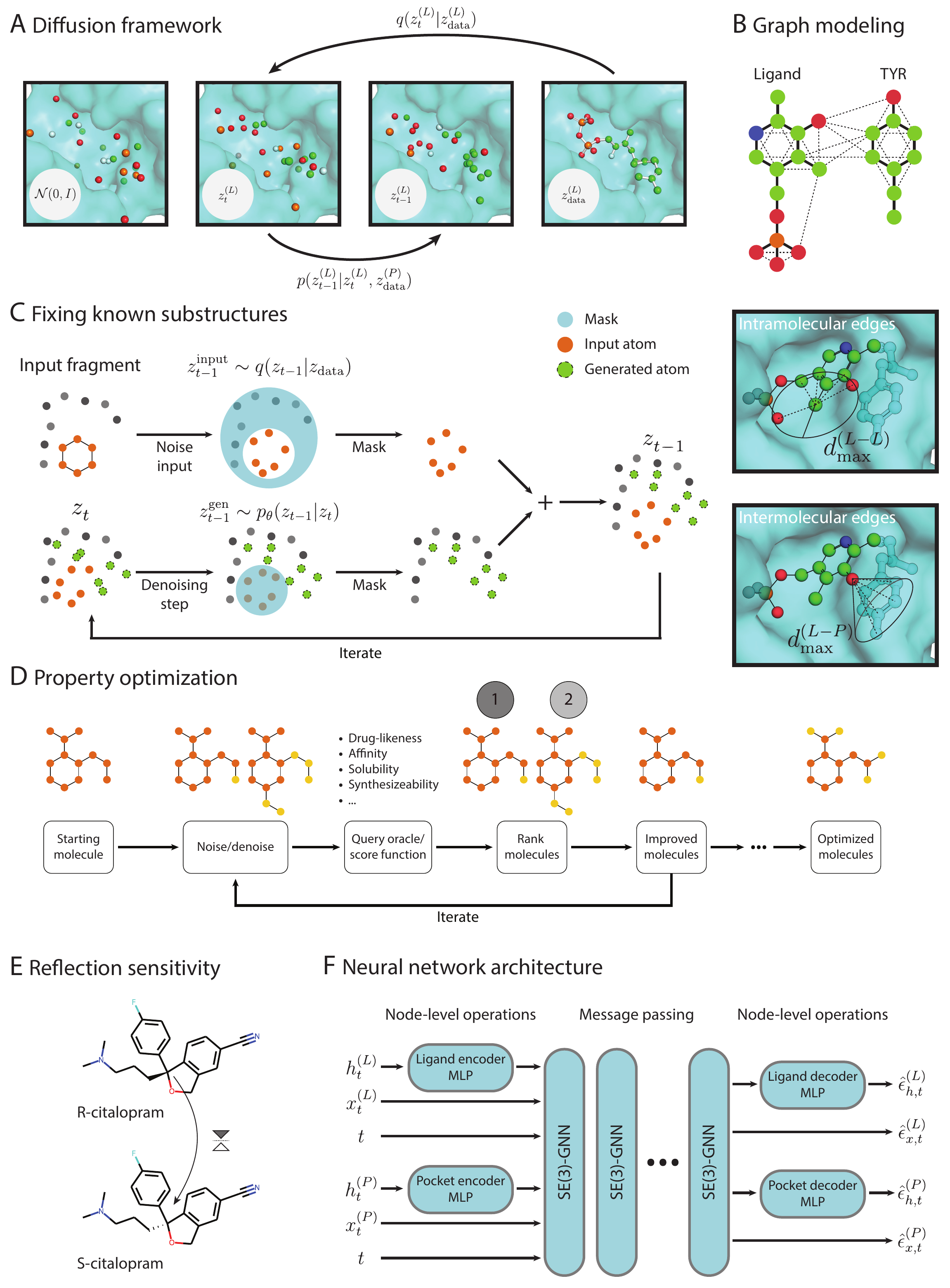}
    \phantomcaption
\end{figure*}
\clearpage

\begin{figure}[h!]
    \ContinuedFloat
    \caption[]{(A) Overview of the 3D diffusion setup. 
    The diffusion process $q$ yields a noised version of the original atomic point cloud for a time step $t\leq T$.
    The neural network model is trained to approximate the reverse process conditioned on the target protein structure $\bm{z}^{(P)}$. 
    Once trained, an initial noisy point cloud is sampled from a Gaussian distribution $\bm{z}_T^{(L)} \sim \mathcal{N}\left( \boldsymbol{0}, \boldsymbol{I}\right)$ and progressively denoised using the learned model.
    Covalent bonds are added to the resultant point cloud at the end of generation. (B) Each state is processed as a graph where edges are introduced according to distance thresholds within the ligand $d_\text{max}^{L-L}$, within the pocket $d_\text{max}^{P-P}$ and between ligand and pocket nodes $d_\text{max}^{L-P}$. 
    (C) Replacement method for fixing molecular substructures. To complete the known part of the molecule (orange) with newly generated chemical matter (green atoms), we apply the learned denoising process to the entire molecule (orange \& green), but at every step we replace the prediction for the known part (orange) with the ground-truth noised version computed with $q$. The protein context (gray) remains unchanged in every step.
    (D) Iterative procedure to tune molecular features. We find variations of a starting molecule by applying small amounts of noise and running an appropriate number of denoising steps. The new set of molecules is ranked by an arbitrary oracle and the procedure is repeated for the strongest candidates. 
    (E) Antidepressant Citalopram as an example in which stereochemistry is essential for its therapeutic effect.
    (F) The neural network backbone is composed of MLPs that map scalar features of ligand and pockets nodes into a joint embedding space, and SE(3)-equivariant message passing layers that operate on these features and each node's coordinates. It outputs the predicted noise values for every vertex.
    }\label{fig:methods}
\end{figure}

\section{Equivariant Diffusion Models for SBDD}
We leverage equivariant denoising diffusion probabilistic models (DDPMs)~\citep{ho2020denoising,hoogeboom2022equivariant} to generate molecules and binding conformations jointly with respect to a specific protein target.
Figure~\ref{fig:methods}A schematically depicts the 3D diffusion procedure. During training, varying amounts of random noise are applied to 3D structures of real ligands and a neural network learns to predict the noise-less features of the molecules. For sampling, these predictions are used to parameterize denoising transition probabilities which allow us to gradually move a sample from a standard normal distribution onto the data manifold.
Both the protein and the ligand are represented as 3D point clouds, where atom types are encoded as one-hot vectors, and all objects are processed as graphs. 
For improved computational efficiency, we define independently tunable distance cutoffs for intermolecular edges between nodes of the ligand and pocket and intramolecular edges between two nodes from the same molecule (Figure~\ref{fig:methods}B). This means information is only propagated between spatially proximal atoms.
Our neural network is designed to respect natural symmetries of the molecular system, which include rotations and translations but excludes nonsuperposable transformations. That is, we process rigid transformations in an equivariant way but not reflections. This design choice is motivated by well-studied examples of drugs whose sterochemistry affects their activity and toxicity. For instance, the antidepressent Citalopram (Figure~\ref{fig:methods}E) has two enantiomers but only the S-enantiomer has the desired therapeutic effect. The difference between the S- and R-form of the molecule, however, is only detectable by a reflection-sensitive GNN (Appendix~\ref{sec:SEGNN}).
To process ligand and pocket nodes with a single graph neural network (GNN), atom types and residue types are first embedded in a joint node embedding space by separate learnable MLPs.
We also experimented with coarse-grained $C_\alpha$ descriptions of the pockets to reduce processing time even further but found this representation to be inferior in most cases (Appendix~\ref{sec:CA_models}).
Further technical details of the diffusion framework and equivariant neural network are described in Method Sections \ref{sec:ddpm} and \ref{sec:EGNN}.

To condition the 3D generative model on the structure of the protein pocket, we consider two distinct approaches. 
In the first approach, DiffSBDD-cond, we provide fixed three-dimensional context in each step of the denoising process. To this end, we supplement the ligand atomic point cloud $\bm{z}^{(L)}_t$, denoted by superscript $L$, with protein pocket nodes $\bm{z}^{(P)}_\text{data}$, denoted by superscript $P$, that remain unchanged throughout the reverse diffusion process (Figure~\ref{fig:methods}A).
For the second method, DiffSBDD-joint, we initially train a diffusion model to approximate the joint distribution $p(\bm{z}^{(L)}_\text{data}, \bm{z}^{(P)}_\text{data})$ of ligand-pocket pairs, and inject information about target pockets only at inference time. The methodology is analogous to the substructure inpainting approach described below (Section~\ref{sec:inpainting} and Figure~\ref{fig:methods}C).
Both approaches are equally applicable to the small molecule design task and in practice differ only in whether the neural networks expects the original pocket or a noisy version as input.

To evaluate our approach we first show that diffusion models are a powerful framework for learning the distribution of three-dimensional molecular data by generating new target-specific ligands \textit{de novo} without additional constraints or optimizing a particular property (Section~\ref{sec:distribution_learning}). We then demonstrate how the flexibility of diffusion models enables partial molecular redesign to incorporate specific design constraints without needing to develop specialised models (Sections~\ref{sec:results_inpainting}), and iterative improvement of molecular properties measured by arbitrary oracles (Section~\ref{sec:optimisation}). While we provide empirical results only for our model, the methodology can be readily used in combination with other recently published diffusion models for molecule design~\citep{guan20233d,lin2022diffbp,xu2023geometric,guan2023decompdiff,weiss2023guided}.
Finally, we have curated an experimentally determined binding dataset derived from Binding MOAD~\cite{hu2005binding}, which supplements the commonly used synthetic dataset CrossDocked~\cite{francoeur2020three}, to validate our model's performance under realistic binding scenarios. 
While protein-ligand pairs of the former might contain non-native contacts, since ligands were cross-docked into structurally similar binding pockets, our new dataset consists exclusively of experimentally validated interactions.

\section{DiffSBDD captures the data distribution faithfully}
\label{sec:distribution_learning}

\begin{table*}[]
    \centering
    \caption{Evaluation of generated molecules for targets from the CrossDocked and Binding MOAD test sets. To assess how well the models capture properties of real ligands, we compute the Wasserstein distance between the distributions of a scores from generated molecules and the ground truth molecules from the test sets. 
    The best performance (i.e. lowest Wasserstein distance) is highlighted in bold.
    $^*$ denotes that we re-evaluate the generated ligands provided by the authors. $^\dag$ means inference times are taken from the original paper. $^\ddag$ means inference time estimated based on five targets. \\
    \emph{QED}: Quantitative Estimation of Drug-likeness~\citep{bickerton2012quantifying}; \emph{SA}: Synthetic Accessibility~\citep{ertl2009estimation}; \emph{LogP}: partition coefficient~\citep{wildman1999prediction}; \emph{CNN}: Convolutional Neural Network.}
    \label{tab:table_distribution_learning}
    \begin{adjustbox}{width=1\textwidth}
    \begin{tabular}{clccccccc}
    \toprule
    & & \multicolumn{6}{c}{Wasserstein distance to reference distribution} & \\
    \cmidrule{3-8}
    & & QED & SA & LogP & Lipinski & Vina score & CNN affinity & Time (s, $\downarrow$) \\
    \midrule
    
    \multirow{4}{*}{\rotatebox[origin=c]{90}{\footnotesize{CrossDocked}}}
    & Pocket2Mol~\cite{peng2022pocket2mol}$^*$ & $0.106$ & $\best{0.24}$ & $0.879$ & $0.647$ & $\best{0.596}$ & $0.616$ & $2504 \pm 2207$$^\dag$ \\
    & ResGen~\citep{zhang2023resgen} & $0.116$ & $0.556$ & $0.778$ & $0.673$ & $2.15$ & $1.21$ & $\approx 936^\ddag$ \\ 
    \cmidrule{2-9}
    & DiffSBDD-cond & $0.0205$ & $1.29$ & $\best{0.589}$ & $\best{0.261}$ & $0.848$ & $\best{0.145}$ & $\best{135.866 \pm 51.66}$ \\
    & DiffSBDD-joint & $\best{0.0187}$ & $1.51$ & $1.67$ & $0.405$ & $0.694$ & $0.281$ & $160.314 \pm 73.30$ \\
    \midrule
    
    \multirow{4}{*}{\rotatebox[origin=c]{90}{\footnotesize{Binding MOAD}}}
    & Pocket2Mol~\cite{peng2022pocket2mol} & $0.123$ & $0.93$ & $0.841$ & $0.269$ & $\best{3.4}$ & $1.58$ & $\approx 613^\ddag$ \\ 
    & ResGen~\citep{zhang2023resgen} & $0.0727$ & $0.92$ & $\best{0.831}$ & $0.237$ & $6.93$ & $1.71$ & $\approx 697^\ddag$ \\
    \cmidrule{2-9}
    & DiffSBDD-cond & $0.0924$ & $1.15$ & $0.833$ & $\best{0.143}$ & $4.26$ & $0.903$ & $\best{336.061 \pm 85.02}$ \\
    & DiffSBDD-joint & $\best{0.0716}$ & $\best{0.76}$ & $1.23$ & $0.183$ & $9.56$ & $\best{0.628}$ & $369.873 \pm 124.54$ \\
    \bottomrule
    \end{tabular}
    \end{adjustbox}
\end{table*}

\begin{figure}
    \centering
    \includegraphics[width=\textwidth]{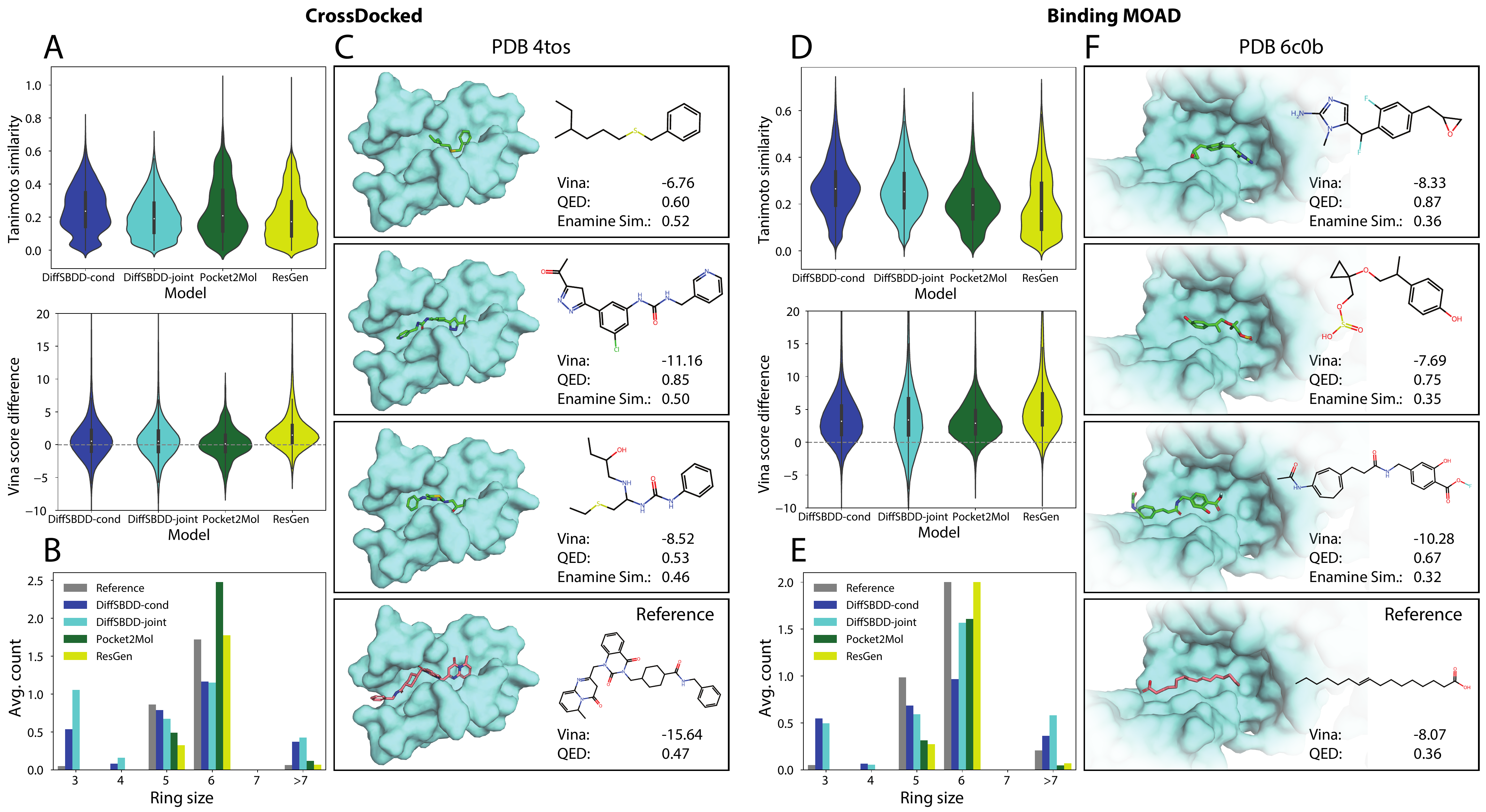}
    \caption{Evaluation of distribution learning capabilities and generated examples. All targets are taken from the CrossDocked and Binding MOAD test sets.
    (A) Comparison of generated molecules with the reference molecule from the same pocket. We compare Tanimoto similarity of the molecular fingerprints and compute the difference $\text{Vina}_\text{gen} - \text{Vina}_\text{ref}$ between their Vina docking scores.
    (B) Average number of rings of different sizes per generated molecule.
    (C) Example molecules generated by DiffSBDD-cond for a pocket from the CrossDocked test set. We compared all generated molecules with the approximately 4.2M compounds from the Enamine Screening Collection, and selected the three closest hits with drug-likeness $\text{QED}>0.5$. Vina docking score, QED drug-likeness score and fingerprint similarity to the most similar Enamine molecules are reported in each case.
    (D-F) Same analyses but for target pockets from the Binding MOAD test set.
    }
    \label{fig:distribution_learning}
\end{figure}

As a first test to our model, we probe its ability to accurately represent the properties of real binders, and compare the results with Pocket2Mol~\citep{peng2022pocket2mol} and ResGen~\citep{zhang2023resgen}, two recently published autoregressive models, which represent the previous state-of-the-art class of machine learning models for structure-based drug design.
We use publicly available code and weights of the models~\citep{pocket2mol-github,resgen-github}.

Since distribution learning capabilities in the high-dimensional space of chemical compounds are difficult to quantify directly, we instead measure a range of molecular properties that are relevant for potential drug candidates. We then compare the distributions of these scores to the distributions we get from the real ligands in our test set using the Wasserstein distance. These results are summarized in Table~\ref{tab:table_distribution_learning} for computational scores of drug-likeness (QED), synthetic accessibility (SA), hydrophobicity (LogP) and two measures of target affinity, the empirical Vina scoring function and a neural network estimation of binding affinity (CNN affinity). Both were computed by GNINA~\citep{mcnutt2021gnina} after local energy minimization to resolve minor clashes.
The underlying distributions of scores are shown in supplementary Figures~\ref{fig:distributions_crossdocked} and \ref{fig:distributions_bindingmoad}.
We perform this analysis both for the test set targets from CrossDocked~\citep{francoeur2020three}, a standard dataset extensively used in prior works~\citep{ragoza2022generating,liu2022generating,peng2022pocket2mol,zhang2023resgen}, and our newly curated dataset based on Binding MOAD~\citep{hu2005binding}.
Note that Pocket2Mol and ResGen were trained on the same CrossDocked training set but not on Binding MOAD.

Generally, our diffusion models capture molecular properties of natural ligands more accurately than the autoregressive baselines despite significantly shorter sampling times. 
A notable exception is the Vina score, which Pocket2Mol matches particularly well on the CrossDocked dataset. Interestingly, this observation is not confirmed by GNINA's CNN affinity which estimates the same quantity. Its distribution is better approximated by DiffSBDD.
Figure~\ref{fig:distribution_learning}A shows that both DiffSBDD and Pocket2Mol Vina scores are centered around the reference but the spread is larger in the case of the diffusion models, which means their samples contain larger fractions of low scoring molecules but also ligands that potentially bind more tightly than the native counterparts.
The greater abundance of high scoring molecules is particularly important in anticipation of downstream design applications, where we often look for the most competitive binder rather than average candidates.
A similar observation holds for the Binding MOAD dataset with experimentally determined binding complexes. However, unlike the CrossDocked case, docking scores are lower on average than the scores of corresponding reference ligands from this dataset. We believe the reason to be twofold: the Binding MOAD training set is much smaller and also contains more challenging ground-truth ligands (native binders) whereas CrossDocked complexes can have unrealistic protein-ligand interactions. This hypothesis is supported by less favorable Vina scores of reference molecules from the synthetic dataset on average (-7.68 vs. -9.17 kcal/mol). 
This result underscores the importance of high quality training sets for SBDD models that aim to design high affinity binders.
Lastly, the DiffSBDD models also produce molecules that are slightly more similar to the reference on average (Figure~\ref{fig:distribution_learning}A,D) and contain a comparable amount of 5- and 6-rings to natural ligands (Figure~\ref{fig:distribution_learning}B,E). However, very small and very large ring systems are typically over-represented in DiffSBDD molecules. 

Panels C and F of Figure~\ref{fig:distribution_learning} present a representative selection of molecules for one target from each test set. 
The selection is filtered to contain examples which are drug-like ($\text{QED}>0.5$) and similar to purchasable molecules from the Enamine Screening Collection. These filters represent favorable properties one might look for in a drug design campaign.
The target with Protein Data Bank (PDB) identifier 6c0b, for example, is a human receptor which is involved in microbial infection~\cite{chen2018structuralfizz} and possibly tumor suppression~\cite{ding2016fzd2}. The reference molecule, a long fatty acid (see Figure~\ref{fig:distribution_learning}F, bottom panel) that aids receptor binding~\cite{chen2018structuralfizz}, has too high a number of rotatable bonds and low a number of hydrogen bond donors/acceptors to be considered a suitable drug (QED of 0.36). Our model however, generates drug-like ($\text{QED}=0.87$ in the first example)
and suitably sized molecules by adding aromatic rings connected by a few rotatable bonds, which allows the molecules to adopt a complementary binding geometry and is entropically favourable by reducing the degrees of freedom, a classic approach in medicinal chemistry~\cite{ritchie2009rings}. 
Larger random samples of generated molecules are presented in the Appendix (Figure~\ref{fig:random_mols}).

\begin{figure}[t!]
    \centering
    \includegraphics[width=0.9\textwidth]{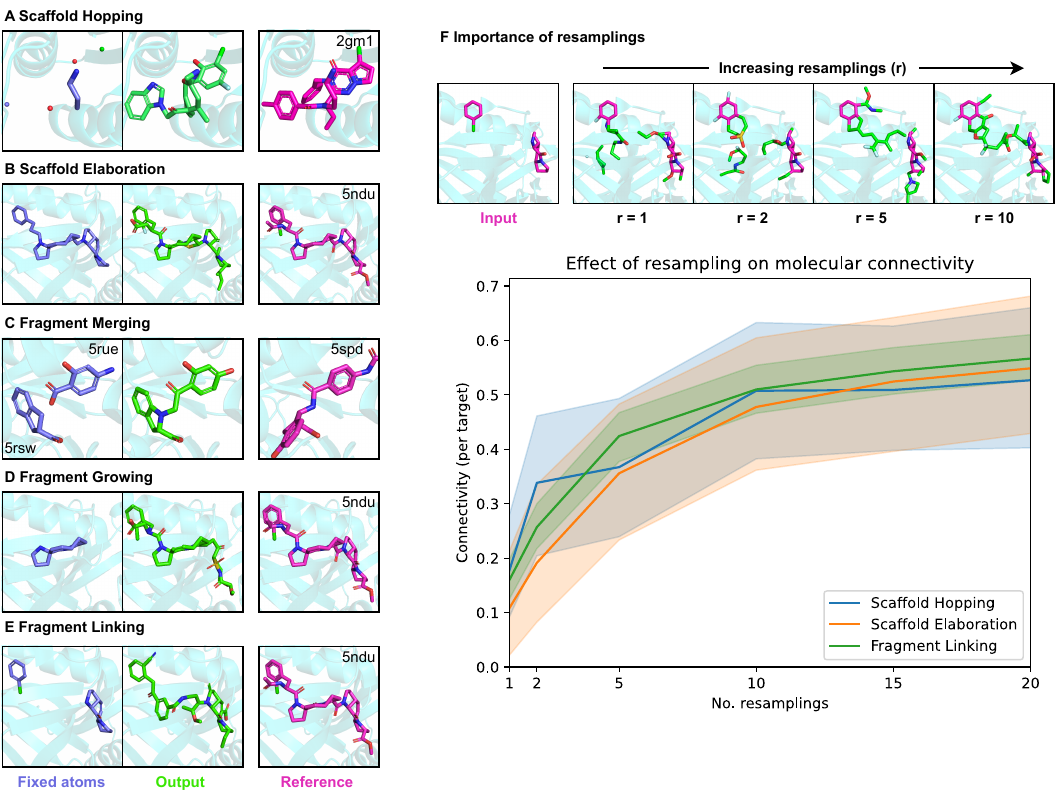}
    \caption{Molecular inpainting design examples for scaffold hopping (A), scaffold elaboration (B), fragment merging (C), fragment growing (D) and fragment linking (E) respectively. The input to our model (the fixed atoms) is shown in blue, the outputs (designed molecules) are shown in green and the original molecule is shown in magenta for reference. PDB codes are shown for the ground truth structure. In the case of fragment merging, we compose fragments with two different crystal structures with PDB codes shown. (F) Importance of resampling for generating realistic and connected molecules. Top: Visual example; inpainted region (green) finally harmonizes with molecular context at high resamplings. Bottom: Effect of the number of resampling steps on molecular connectivity. Means and 95\% confidence intervals are plotted for 3 design tasks. For this experiment we used 20 randomly selected targets from the test set.}
    \label{fig:inpainting_redesign}
\end{figure}

\section{Generating novel chemical matter from known substructures}
\label{sec:results_inpainting}

In drug discovery it is very common to design molecules around previously identified potent substructures. For example, we may wish to design a scaffold around a set of functional groups (scaffold hopping) or extend an existing fragment to make a whole molecule (fragment growing). 
Generating compounds, or parts thereof, conditioned on a given molecular context is reminiscent of inpainting, a technique originally introduced for completing missing parts of images~\cite{song2020score, lugmayr2022repaint} but also adopted in other domains, including biomolecular structures~\cite{wang2022scaffolding}.
We can realise a number of drug discovery sub-tasks via an inpainting technique known as the `replacement method'~\cite{ho2022video,lugmayr2022repaint}, whereby we add new atoms in and around fixed regions of the substructure to design whole molecules (Fig. \ref{fig:methods}C and Methods Section~\ref{sec:inpainting}). Unlike previous methods, using DiffSBDD in this way does not require retraining a model on any specialized or synthetic datasets. 
Curating such datasets is often time and labor-intensive, and typically relies on potentially sub-optimal assumptions (e.g. definition of fragments) to convert a general dataset of small molecules into a task-specific dataset that can be used to train specialised models.
With our proposed approach, by contrast, the simple definition of an arbitrary binary mask is sufficient for the diffusion model to generalize to any inpainting task whilst using a neural network trained on all available protein-ligand data in raw form.

\begin{table*}[t]
    \centering
    \caption{Evaluation of molecular inpainting for fragment linking, scaffold hopping and scaffold elaboration across the whole Binding MOAD test set. Percentage value next to task name denotes the proportion of atoms fixed during the design process. DiffSBDD-\textit{baseline} generates a whole new molecule from scratch, DiffSBDD-\textit{diversify} is elaborating around a fixed substructure of an existing molecule and DiffSBDD-\textit{de novo} is designing new motifs around a fixed substructure from stratch. $t$ and $r$ are the number of partial noising steps or resamplings for \textit{diversify} and \textit{de novo} respectively.
    Note, here we use a subset of our original test set for which we can calculate masks for all design tasks ($n=55$). 
    }
    \label{tab:main_table}
    \begin{adjustbox}{width=1\textwidth}
    \begin{tabular}{lcccccc}
    \toprule
    & Vina ($\downarrow$) & Validity ($\uparrow$) & Connectivity ($\uparrow$) & QED ($\uparrow$) & SA ($\downarrow$) & Diversity ($\uparrow$)\\ 
    \midrule
    Test set & $-9.86 \pm 1.6$ & $1$ & $1$ & $0.543 \pm 0.16$ & $3.86 \pm 1.2$ & --- \\ 
    DiffSBDD-\textit{baseline} & $-5.69 \pm 6$ & $0.964$ & $0.589$ & $0.419 \pm 0.2$ & $4.99 \pm 1.1$ & $0.701 \pm 0.09$ \\ 
    \midrule
    \textbf{Fragment linking (73.43\% atoms fixed)}\\
    \midrule
    DiffSBDD-\textit{de novo} ($r=20$) & $-7.74 \pm 1.8$ & $0.796$ & $0.558$ & $0.322 \pm 0.17$ & $5.38 \pm 0.76$ & $0.486 \pm 0.09$ \\
    DiffSBDD-\textit{diversify} ($t=100$) & $-8.72 \pm 1.7$ & $0.991$ & $0.968$ & $0.476 \pm 0.14$ & $4.26 \pm 1.1$ & $0.35 \pm 0.089$ \\
    
    DiffLinker \cite{igashov2022equivariant} & $-6.92 \pm 2.7$ & $0.947$ & $0.840$ & $0.349 \pm 0.19$ & $4.72 \pm 0.97$ & $0.453 \pm 0.13$ \\ 

    \midrule
    \textbf{Scaffold hopping (27.32\% atoms fixed)} \\
    \midrule
    DiffSBDD-\textit{de novo} ($r=20$) & $-7.6 \pm 2.5$ & $0.782$ & $0.663$ & $0.39 \pm 0.18$ & $5.29 \pm 0.72$ & $0.612 \pm 0.074$ \\ 
    DiffSBDD-\textit{diversify} ($t=100$) & $-8.95 \pm 1.8$ & $0.977$ & $0.948$ & $0.492 \pm 0.18$ & $4.39 \pm 1.0$ & $0.479 \pm 0.1$ \\

    \midrule
    \textbf{Scaffold elaboration (72.68\% atoms fixed)}\\
    \midrule
    DiffSBDD-\textit{de novo} ($r=20$) & $-8.1 \pm 1.8$ & $0.852$ & $0.445$ & $0.388 \pm 0.2$ & $5.2 \pm 0.7$ & $0.397 \pm 0.11$ \\ 
    DiffSBDD-\textit{diversify} ($t=100$) & $-9.32 \pm 1.7$ & $0.995$ & $0.971$ & $0.516 \pm 0.18$ & $4.12 \pm 1.1$ & $0.282 \pm 0.14$ \\

    \bottomrule
    \end{tabular}
    \end{adjustbox}
\end{table*}

Examples of molecules designed with inpainting for scaffold hopping, scaffold elaboration, fragment merging, fragment growing, and fragment linking are shown in Figure \ref{fig:inpainting_redesign}. All molecular inpainting experiments use a version of DiffSBDD trained on Binding MOAD.
Figure \ref{fig:inpainting_redesign}A shows an example of scaffold hopping a design for a mitotic kinesin Eg5 inhibitor (PDB code 2gm1) \cite{kim2006synthesis} where we fix the functional groups mediating the binding to the pocket whilst designing a new scaffold structure. 
Figure \ref{fig:inpainting_redesign}B shows the opposite case of scaffold elaboration for a rationally designed oncology inhibitor targeting a phosphoprotein (PDB code 5ndu) \citep{barone2020designed} where we fix the scaffold and design new functional groups. 
Figure \ref{fig:inpainting_redesign}C shows an example of fragment merging. We successfully replicate the results of \citet{gahbauer2023fragment_merge}, which performed fragment merging of two fragments (PDB code 5rsw and 5rue) identified by experimental screening \cite{schuller2021fragment_screen} for the SARS-CoV-2 non-structural protein 3 (Nsp3) using the chemoinformatics-based method Fragmenstein\footnote{www.github.com/matteoferla/Fragmenstein}. 
Figure \ref{fig:inpainting_redesign}D shows an example of fragment growing around the central motif of PDB entry 5ndu.
Finally, Figure \ref{fig:inpainting_redesign}E gives an example of fragment linking between two outer fragments of PDB entry 5ndu. Note that here we are not only designing a small linker made of a few atoms but rather an entirely new fragment with two connecting linkers. Such a linker design would be substantially out-of-distribution for previous methods, due to the size and complexity of the linker relative to those used during training \cite{imrie2020delinker}.
Further implementation details of the fragment merging experiment are given in Appendix Section \ref{app:fragment_merging}.

Depending on the use case, we find it desirable to perform molecular inpainting within two regimes; (i) designing a completely new inpainted region \textit{de novo} (DiffSBDD-\textit{de novo}) as to explore the entire chemical fitness landscape, or (ii) redesigning an existing region (e.g. a scaffold) via partial noising then denoising (see Section \ref{supp:optimisation}), thus locally exploring desired properties by exploitation (DiffSBDD-\textit{diversify}).  The first case is more amenable to situations in which we have \emph{no prior information} other than the fixed substructure (e.g. fragment linking after a fragment screen), meaning that unconstrained exploration of the chemical fitness landscape is the preferred approach for the majority of SBDD. The second case is more relevant in scenarios where we \emph{have prior information} about the desired chemical and topological composition of the designed region which we can use to bias generation (with the choice of $t$ being a hyperparameter). This is particularly relevant in the case of scaffold hopping, where we are trying to keep the properties of a molecule relatively unchanged whilst designing a new topology \cite{bohm2004scaffoldhopping}.
To investigate the differences between both approaches quantitatively,
we further tested our method systematically for the whole Binding MOAD test set in the tasks of linker design, scaffold hopping and scaffold elaboration (Table \ref{tab:main_table}). As baselines, we provide the metrics for the molecules in the test set, molecules only conditioned on the pocket with no fixed substructure (DiffSBDD-\textit{baseline}) and the performance of DiffLinker \cite{igashov2022equivariant} in the fragment linking task.
Due to the fact that we can only calculate fragment and scaffold masks for larger molecules, we have reduced the size of the test set used in Table~\ref{tab:main_table} to $n=55$ to ensure fair comparison between methods. 

Constraining fixed regions to highly complementary substructures within the protein pocket significantly enhances Vina scores using DiffSBDD-\textit{de novo} compared to the DiffSBDD-\textit{baseline}.
For molecules generated without substructure conditioning under DiffSBDD-\textit{baseline}, the average Vina score is -5.69. In contrast, DiffSBDD-\textit{de novo} sees improved scores, achieving -7.74 kcal/mol when used for linker design from starting fragments and achieves results comparable to the specialist model DiffLinker. In the case of scaffold elaboration, DiffSBDD-\textit{de novo} significantly boosts docking scores from -5.69 kcal/mol to -8.10 kcal/mol over the baseline, by focusing on the optimal placement of functional groups to facilitate key residue binding on a pre-existing scaffold.
Moreover, there is a notable improvement in average docking scores for scaffold hopping, with scores rising from -5.69 kcal/mol to -7.60 kcal/mol when using DiffSBDD-\textit{de novo}. This enhancement is achieved despite a relatively small proportion of atoms being fixed, about 27.32\%, compared to other tasks. The significant increase in docking scores is attributed to the nature of the fixed atoms, primarily functional groups that form the pharmacophore, which are crucial for binding affinity.
For detailed implementation, see Appendix~\ref{supp:quantitative_inpainting}.

Similar to earlier findings, the replacement method often produces poor and inconsistent outcomes. Following the approach by \citet{lugmayr2022repaint}, we enhanced the sample quality by refining intermediate states iteratively before progressing in the denoising process, a technique called resampling, as detailed in the Methods Section~\ref{sec:inpainting}. This technique is crucial for seamlessly integrating modified and original areas and proves essential in the molecular context. Minimal resampling resulted in chemically valid but disjointed structures, while increased iterations led to coherent molecules, even in complex scenarios with extensive modifications needed (Fig.~\ref{fig:inpainting_redesign}F). Our results indicate that the effect of resampling on molecular connectivity is particularly pronounced (Fig.~\ref{fig:inpainting_redesign}F) but it also significantly impacts another metrics (Appendix Figure \ref{fig:resampling_metrics}B-D). 

\section{Molecule optimization: iterative search for better molecule candidates}\label{sec:optimisation}

For hit identification and optimization of lead molecules in real use cases, it is not enough to just sample molecules from the whole training data distribution. Instead, we are usually interested in the better performing tail of the distribution, and only want to pursue the most promising candidates. Since we could show that DiffSBDD recapitulates the chemical space of the training set including high-scoring molecules, we should always find promising drug candidates with strong docking scores, synthetic accessibility and other desired properties. Here we propose a simple protocol to access them efficiently (Figure \ref{fig:methods}D).

We first noise a molecule from an experimental protein-ligand complex for $t$ steps, where $t \ll T$, using the forward diffusion process. From this partially noised sample, we can then denoise the appropriate number of steps with the reverse process until $t=0$. The stochasticity in this quick noise/denoise process allows us to sample new and diverse candidates of various properties whilst staying in the same region of chemical space, assuming $t$ is small (see Appendix Figure \ref{fig:noise_denoise}). This approach is inspired by~\cite{luo2022antigen} but note this does not allow for direct optimization of specific properties. Instead, it can be regarded as an exploration around the local chemical space whilst maintaining high shape and chemical complementary via the conditional denoising model.

We extend this idea by combining the partial noising/denoising procedure with a simple evolutionary algorithm that optimizes for specific molecular properties (Figure~\ref{fig:methods}D). 
We find that our model performs well at this task out of the box without requiring additional fine-tuning. At every stage in the optimization process, we generate 100 new molecules (from either the previous generation or the original molecule in the first case). Molecules are modified via partial noising/denoising with a randomly chosen $t$ between 10 and 150. The new molecules are then passed to an oracle/score function (e.g. docking program or synthetic accessibility predictor) to be ranked. The top $k$ molecules are then selected to seed the new population. In our study, we use $k=10$.

\begin{figure}[t!]
    \centering
    \includegraphics[width=\textwidth]{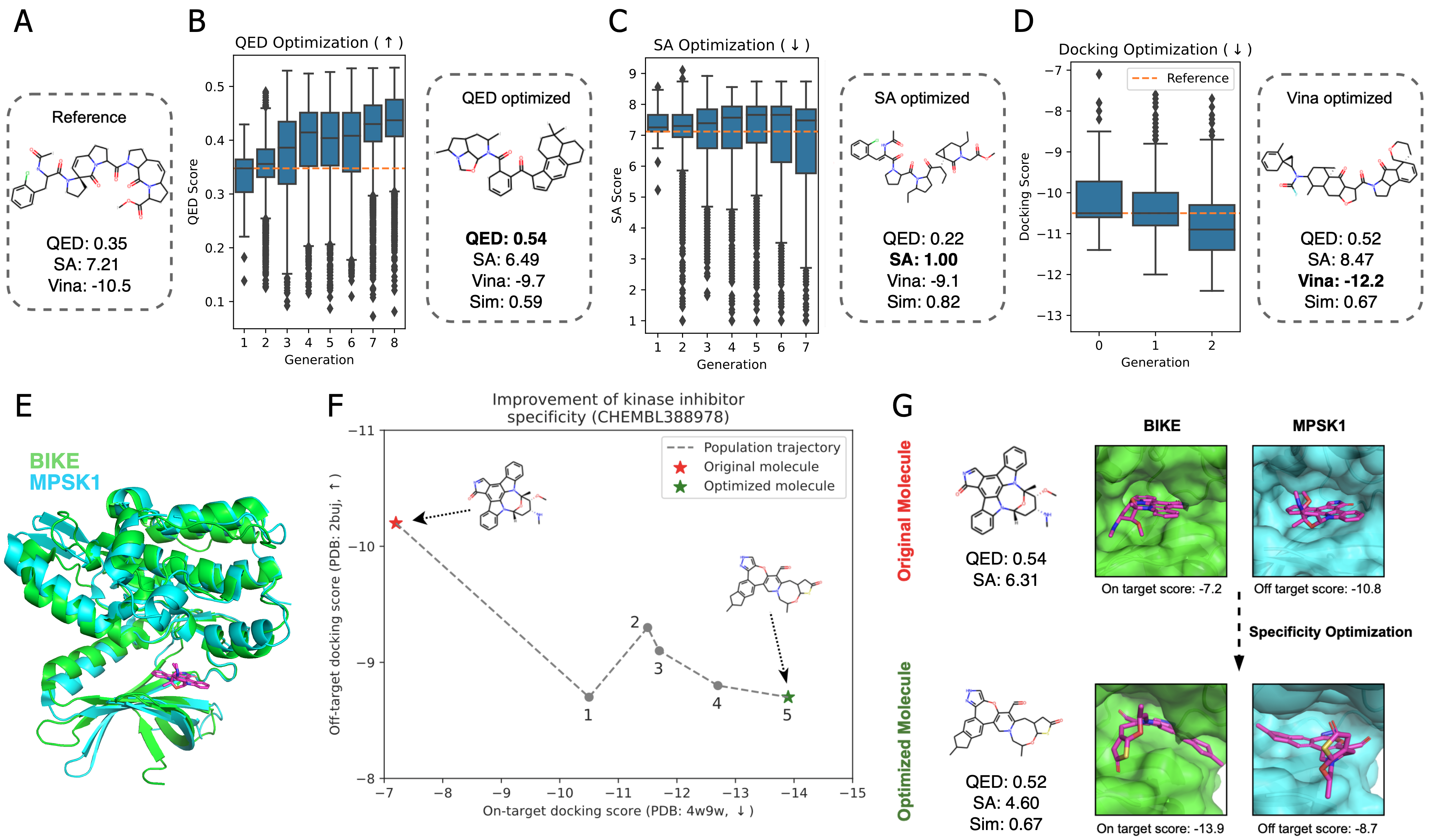}
    \caption{Results on molecular optimization using DiffSBDD. (A-D) Experiments on single property molecular optimization. (A) Starting molecule from PDB code 5NDU. (B) QED optimization over 8 generations. (C) SA optimization over 7 generations. (D) docking score optimization over 3 generations. We found that optimization over subsequent generations continuously optimized the docking score, but that was at expense of molecular quality. 
    (E-G) Kinase inhibitor specificity optimization experiment. (E) Cartoon representation showing the high degree of structural similarity between our two kinases of interest (BIKE and MPSK1). (F) Trajectory plot showing the highest scoring molecule at each iteration during kinase inhibitor optimization. (G) Visual representation of the molecular graphs and bound conformations of the native and final molecules with corresponding Vina docking scores.
    Boxes in panels (B-D) represent the upper and lower quartile as well as the median of the data. Whiskers denote 1.5 times the interquartile range. Outliers outside this range are shown as flier points.
    }
    \label{fig:optimisation}
\end{figure}

In Figure~\ref{fig:optimisation}A-D, we optimize an inhibitor molecule found in PDB entry 5ndu~\cite{barone2020designed}, which has poor SA and QED scores, 7.21 and 0.35 respectively, but high binding affinity. Over a number of rounds of optimization, we can observe substantial increases in QED, from 0.35 to mean of 0.43, whilst still maintaining high similarity to the original molecule. We can also rescue the low synthetic accessibility score of the seed molecule by producing a battery of highly accessible molecules when selecting for SA. Finally, we observe that we can perform significant optimization of binding affinity after only a few rounds of optimization.

To demonstrate the power of molecular optimization with DiffSBDD, we consider the challenging case of \emph{highly selective kinase inhibitor design} (Figure \ref{fig:optimisation}E-G).
In our experiment, we perform positive design against our on-target kinase BIKE (PDB code 4w9w) \cite{sorrell2016kinase_selective2} whilst simultaneously performing negative design against the structurally similar off-target kinase MPSK1 (PDB code 2buj) \cite{debreczeni20052buj} (Fig.~\ref{fig:optimisation}E) by optimizing for $L$. Furthermore, before selecting the top molecules, we pruned any candidates that regress with regard to the on- and off-target docking scores of the original molecule (i.e. those above or left of the red star in Fig.~\ref{fig:optimisation}F) 
in order to bias the molecules to have high affinity to the on-target kinase as well as specificity. The starting molecule (ChEMBL identifier CHEMBL388978) had an on- and off-target docking score of \num{-7.2} kcal/mol and \num{-10.8} kcal/mol respectively. After 5 rounds of optimization, we had improved the on- and off-target docking scores for the top molecule to \num{-13.9} kcal/mol and \num{-8.7} kcal/mol respectively (Fig \ref{fig:optimisation}G), demonstrating substantially improved specificity and reduced off-target activity. Furthermore, all of this is achieved whilst maintaining a high QED score (from 0.54 to 0.52) and improving the SA score from 6.31 to 4.60.

\section{Conclusion}\label{sec:conclusion}

Many machine learning methods for structure-based drug design create molecules one atom at a time, thereby imposing an arbitrary order on the generative process and preventing them from reasoning about molecules holistically. Moreover, most existing approaches focus exclusively on the \textit{de novo} generation of new ligands from scratch, which often limits their sample quality and synthesizability, and ultimately hinders lab validation of designs. 
In this work, we investigated the distribution learning capabilities of 3D-conditional diffusion models as an alternative to the autoregressive paradigm. To this end, we developed DiffSBDD, an $SE(3)$-equivariant 3D-conditional diffusion model that generates small molecule ligands for given target pocket structures, with chemical properties that closely match those of native ligands.
Since, on the application side, medicinal chemists typically have concrete design specifications in mind, we studied different substructure redesign and optimization techniques, and showed how these can be used to improve a range of desirable properties of candidate compounds, such as computational docking and drug-likeness scores.
Constraining the problem to realistic substructures like fragments or scaffolds leads to better designs because it prevents the neural network from overly hallucinating.
Retaining substructures of previously synthesized molecules facilitates chemical synthesis and experimental testing. 
Moreover, the capability to further `locally' optimize designed ligands is important in real-world drug discovery and effectively improves the quality of the initial designs.
For similar applications, previous works typically resorted to specialised models that were trained on tailored datasets and performed well only on narrowly defined tasks. Here, we provided evidence that a powerful general diffusion model can be used as a drop-in replacement for these specialised models if the sampling procedure is modified appropriately.
This means we can expect better performance in all discussed sub-tasks, solely by improving the distribution learning capabilities and sample quality of the main model. 

In general, data-driven techniques are emerging as a suitable tool to tackle the immense complexity of the biochemical design space, which is extremely hard to navigate with mechanistic approaches.
While the purely \textit{de novo} design of novel chemical matter remains challenging, we could show that learning-based tools are ready to be incorporated in drug development pipelines if additional design constraints are enforced.
In the future, we foresee an increasing importance of machine learning models in this domain as they are improved further. 
Developing more informative metrics and more reliable benchmarks will be an important step towards reaching this goal because they reduce the reliance on visual inspection of examples and expert judgment and thus accelerate progress.
Lastly, we envision that other iterative sampling techniques, including flow matching and Schr\"{o}dinger bridge models, will benefit from the presented inference strategies as well.

\section{Methods}\label{sec:methods}

\subsection{Denoising Diffusion Probabilistic Models}\label{sec:ddpm}

Denoising diffusion probabilistic models (DDPMs)~\cite{sohl2015deep, ho2020denoising} are a class of generative models inspired by non-equilibrium thermodynamics. Briefly, they define a Markovian chain of random diffusion steps by slowly adding noise to sample data and then learning the reverse of this process (typically via a neural network) to reconstruct data samples from noise.

In this work, we closely follow the framework developed by \citet{hoogeboom2022equivariant}. In our setting, data samples are atomic point clouds $\bm{z}_{\text{data}}=[\bm{x}, \bm{h}]$ with 3D geometric coordinates $\bm{x}\in\mathbb{R}^{N \times 3}$ and categorical features  $\bm{h}\in\mathbb{R}^{N \times d}$, where $N$ is the number of atoms. A fixed noise process
\begin{equation}\label{equ:forward_noise_process}
    q(\bm{z}_t | \bm{z}_\text{data}) = \mathcal{N}(\bm{z}_t | \alpha_t \bm{z}_\text{data}, \sigma_t^2 \bm{I})
\end{equation}
adds noise to the data $\bm{z}_\text{data}$ and produces a latent noised representation $\bm{z}_t$ for $t=0,\dotsc,T$. $\alpha_t$ controls the signal-to-noise ratio $\text{SNR}(t)=\alpha_t^2/\sigma_t^2$ and follows either a learned or pre-defined schedule from $\alpha_0\approx 1$ to $\alpha_T\approx0$ \cite{kingma2021variational}. We choose a variance-preserving noising process~\cite{song2020score} with $\alpha_t = \sqrt{1 - \sigma_t^2}$. 

Since the noising process is Markovian, we can write the denoising transition from time step $t$ to $s < t$ in closed form as
\begin{equation}
    q(\bm{z_s}|\bm{z}_\text{data},\bm{z}_t)=\mathcal{N}\Big(\bm{z}_s\big|\frac{\alpha_{t|s}\sigma_s^2}{\sigma_t^2}\bm{z}_t+\frac{\alpha_s \sigma_{t|s}^2}{\sigma_t^2}\bm{z}_\text{data}, \frac{\sigma_{t|s}^2\sigma_s^2}{\sigma_t^2}\bm{I}\Big)
\end{equation}
with $\alpha_{t|s}=\frac{\alpha_t}{\alpha_s}$ and $\sigma_{t|s}^2=\sigma_t^2 - \alpha_{t|s}^2\sigma_s^2$ following the notation of \citet{hoogeboom2022equivariant}. This true denoising process depends on the data sample $\bm{z}_{\text{data}}$, which is not available when using the model for generating new samples. Instead, a neural network $\phi_{\theta}$ is used to approximate the sample $\hat{\bm{z}}_\text{data}$. More specifically, we can reparameterize Equation~\eqref{equ:forward_noise_process} as $\bm{z}_t=\alpha_t\bm{z}_\text{data}+\sigma_t\bm{\epsilon}$ with $\bm{\epsilon}\sim \mathcal{N}(\bm{0}, \bm{I})$ and directly predict the Gaussian noise $\hat{\bm{\epsilon}}_\theta=\phi_\theta(\bm{z}_t,t)$. Thus, $\hat{\bm{z}}_\text{data}$ is simply given as $\hat{\bm{z}}_\text{data}=\frac{1}{\alpha_t}\bm{z}_t - \frac{\sigma_t}{\alpha_t}\hat{\bm{\epsilon}}_\theta$.

The neural network is trained to maximize the likelihood of observed data by optimizing a variational lower bound on the data, which is equivalent to the simplified training objective~\cite{ho2020denoising,kingma2021variational} $\mathcal{L}_\text{train} = \frac{1}{2}||\bm{\epsilon}-\phi_{\theta}(\bm{z}_t,t)||^2$ up to a scale factor (see Appendix~\ref{supp:ELBO} for details).

\subsection{Equivariance}

Structural biology remains a rather data-sparse domain. It is therefore common practice to encode known geometric constraints, typically equivariance to rotations and translations, directly into the neural network architecture, thereby facilitating the learning task because possible neural operations are limited to a meaningful subset.
In the 3D molecule generation setting, we explicitly exclude reflection-equivariant operations because they would make the model blind to some aspects of stereochemistry. It is known that different stereoisomers can have significantly different therapeutic effects (e.g., \citep{lepola2004equivalent}, Figure~\ref{fig:methods}E) and might even lead to unforeseen off-target activity and hence toxicity.
We therefore developed a reflection-sensitive system that is $\underline{S}E(3)$- rather than $E(3)$-equivariant although the latter is more commonly adopted in related works~\citep{satorras2021n,hoogeboom2022equivariant,guan20233d}.

Technically, we ensure $SE(3)$-equivariance in the following sense\footnote{We ignore node type features, which transform invariantly, for simpler notation.}: 
evaluating the likelihood of a molecule $\bm{x}^{(L)}\in\mathbb{R}^{3 \times N_L}$ given the three-dimensional representation of a protein pocket $\bm{x}^{(P)}\in\mathbb{R}^{3 \times N_P}$ should not depend on global $SE(3)$-transformations of the system, i.e. $p(\bm{R}\bm{x}^{(L)}+\bm{t}|\bm{R}\bm{x}^{(P)}+\bm{t}) = p(\bm{x}^{(L)}|\bm{x}^{(P)})$ for orthogonal $\bm{R}\in\mathbb{R}^{3\times 3}$ with $\bm{R}^T\bm{R}=\bm{I}$, $\det(\bm{R})=1$ and $\bm{t}\in\mathbb{R}^3$ added column-wise. At the same time, it should be possible to generate samples $\bm{x}^{(L)}\sim p(\bm{x}^{(L)}|\bm{x}^{(P)})$ from this conditional probability distribution so that equivalently transformed ligands $\bm{R}\bm{x}^{(L)}+\bm{t}$ are sampled with the same probability if the input pocket is rotated and translated and we sample from $p(\bm{R}\bm{x}^{(L)}+\bm{t}|\bm{R}\bm{x}^{(P)}+\bm{t})$. 
This definition explicitly excludes reflections which are connected with chirality and can alter the biomolecule's properties.

In our set-up, equivariance to the orthogonal group $O(3)$ (comprising rotations and reflections) is achieved because we model both prior and transition probabilities with isotropic Gaussians where the mean vector transforms equivariantly with respect to rotations of the context (see ~\citet{hoogeboom2022equivariant} and Appendix~\ref{sec:proofs}). Ensuring translation equivariance, however, is harder  because the transition probabilities $p(\bm{z}_{t-1}|\bm{z}_t)$ are not inherently translation-equivariant. 
In order to circumvent this issue, we follow previous works~\cite{kohler2020equivariant,xu2022geodiff, hoogeboom2022equivariant} by  limiting the whole sampling process to a linear subspace where the center of mass (CoM) of the system is zero. In practice, this is achieved by subtracting the center of mass of the system before performing likelihood computations or denoising steps. 
Since equivariance of the transition probabilities depends on the parameterization of the noise predictor $\hat{\bm{\epsilon}}_{\theta}$, we can make the model sensitive to reflections with a simple additive cross-product term in the EGNN's coordinate update as discussed in Section~\ref{sec:EGNN} and Appendix~\ref{sec:SEGNN}.

\subsection{$SE(3)$-equivariant Graph Neural Networks}\label{sec:EGNN}
A function $f: \mathcal{X} \rightarrow \mathcal{Y}$ is said to be {\em equivariant} w.r.t. the
group $G$ if $f(g.\boldsymbol{x}) = g.f(\boldsymbol{x})$, where $g.$ denotes the action of the group element $g \in G$ on $\mathcal{X}$ and $\mathcal{Y}$ \cite{serre1977linear}. Graph Neural Networks (GNNs) are learnable functions that process graph-structured data in a permutation-equivariant way, making them particularly useful for molecular systems where nodes do not have an intrinsic order. 
Permutation invariance means that 
$\mathrm{GNN}(\boldsymbol{\Pi}\mathbf{X}) = \boldsymbol{\Pi}\, \mathrm{GNN}(\mathbf{X})$ where $\boldsymbol{\Pi}$ is an $n\times n$ permutation matrix acting on the node feature matrix.

Since the nodes of the molecular graph represent the 3D coordinates of atoms, we are interested in additional equivariance w.r.t. the Euclidean group $E(3)$ or rigid transformations. An $E(3)$-equivariant GNN (EGNN) satisfies 
$\mathrm{EGNN}(\boldsymbol{\Pi}\mathbf{X}\mathbf{A} + \mathbf{b}) = \boldsymbol{\Pi}\, \mathrm{EGNN}(\mathbf{X})\mathbf{A} + \mathbf{b}$ for an orthogonal $3\times 3$ matrix $\mathbf{A}$ with $\mathbf{A}^\top \mathbf{A} = \mathbf{I}$ and some translation vector $\mathbf{b}$ added row-wise.

In our case, since the nodes have both geometric atomic coordinates $\boldsymbol{x}$  as well as atomic type features $\boldsymbol{h}$, we can use a simple implementation of EGNN proposed by \citet{satorras2021n}, in which the updates for features $\bm{h}$ and coordinates $\bm{x}$ of node $i$ at layer $l$ are computed as follows:
\begin{align}
    \boldsymbol{m}_{ij} &= \phi_e(\boldsymbol{h}^l_i, \boldsymbol{h}^l_j, d_{ij}^2, a_{ij}), ~
    \tilde{e}_{ij} = \phi_\text{att}(\bm{m}_{ij}) \label{egnn_message}\\
    \boldsymbol{h}_{i}^{l+1} &= \phi_h(\boldsymbol{h}^l_i, \sum_{j\neq i}\tilde{e}_{ij}\boldsymbol{m}_{ij}) \label{egnn_node} \\
    \boldsymbol{x}_{i}^{l+1} &= \boldsymbol{x}_{i}^{l} + \sum_{j\neq i} \frac{\boldsymbol{x}_{i}^{l} - \boldsymbol{x}_{j}^{l}}{d_{ij} + 1}  \phi_x(\boldsymbol{h}^l_i, \boldsymbol{h}^l_j, d_{ij}^2, a_{ij}) \label{egnn_coordinate}
\end{align}

where $\phi_e$, $\phi_\text{att}$, $\phi_h$ and $\phi_h$ are learnable Multi-layer Perceptrons (MLPs) and $d_{ij}$ and $a_{ij}$ are the relative distances and edge features between nodes $i$ and $j$ respectively. 
Following \cite{igashov2022equivariant}, we do not update the coordinates of nodes that belong to the pocket to ensure the three-dimensional protein context remains fixed throughout the EGNN layers.

We can break the symmetry to reflections and thereby make the GNN layer $\underline{S}E(3)$-equivariant by adding a cross product-dependent term to the coordinate update, which changes sign under reflection:
\begin{align}
    \label{equ:se3gnn_coord}
    \bm{x}_{i}^{l+1} = \bm{x}_{i}^{l} &+ \sum_{j\neq i} \frac{\bm{x}_{i}^{l} - \bm{x}_{j}^{l}}{d_{ij} + 1} \phi_x^d(\bm{h}^l_i, \bm{h}^l_j, d_{ij}^2, a_{ij}) \\ &+ \frac{(\bm{x}_{i}^{l} - \bar{\bm{x}}^l) \times (\bm{x}_{j}^{l} - \bar{\bm{x}}^l)}{||(\bm{x}_{i}^{l} - \bar{\bm{x}}^l) \times (\bm{x}_{j}^{l} - \bar{\bm{x}}^l)|| + 1} \phi_x^\times(\bm{h}^l_i, \bm{h}^l_j, d_{ij}^2, a_{ij}).
\end{align}
Here, $\bar{\bm{x}}^l$ denotes the center of mass of all nodes at layer $l$. $\phi_x^\times$ is an additional MLP. This modification is discussed in more detail in Appendix~\ref{sec:SEGNN}.

\subsection{Inpainting}\label{sec:inpainting}

For molecular inpainting as shown in Figure~\ref{fig:methods}C, a subset of all atoms is fixed and serves as the molecular context we want to condition on. All other atoms are generated by the DDPM. To this end, we diffuse the fixed atoms at each time step and predict a new latent representation $z_{t-1}^\text{gen}$ with the neural network. We then replace the generated atoms corresponding to fixed nodes with their forward noised counterparts:
\begin{align}
    \bm{z}_{t-1}^\text{input} &\sim q(\bm{z}_{t-1}|\bm{z}_\text{data}) \label{equ:inpaint_known} \\
    \bm{z}_{t-1}^\text{gen} &\sim p_\theta(\bm{z}_{t-1}|\bm{z}_{t}) \label{equ:inpainting_transition} \\
    \bm{z}_{t-1} &= \big[ \bm{z}_{t-1}^\text{input}, \bm{z}_{t-1, i\notin\mathcal{M}}^\text{gen} \big], \label{equ:inpaint_combined}
\end{align}
where $\mathcal{M}$ denotes the set of mask indices used to uniquely identify nodes corresponding to fixed atoms.
In this manner, we traverse the Markov chain in reverse order from $t=T$ to $t=0$ to generate conditional samples. Because the noise schedule decreases the noising process's variance to almost zero at $t=0$ (Equation~\eqref{equ:forward_noise_process}), the final sample is guaranteed to contain an unperturbed representation of the fixed atoms.
This approach can be applied to pocket-conditioned ligand-inpainting by fixing all pocket nodes when sampling from the joint distribution model. However, it is much more general and allows us to mask and replace arbitrary parts of the ligand-pocket system without retraining, and can also be combined with the conditionally trained model---an option we explore in Section~\ref{sec:results_inpainting}.

\paragraph{Equivariance}

Since the equivariant diffusion process is defined for a CoM-free system, we must ensure that this requirement remains satisfied after the substitution step in Equation~\eqref{equ:inpaint_combined}. To prevent a CoM shift, we therefore translate the fixed atom representation so that its center of mass coincides with the predicted representation: $\tilde{\bm{x}}_{t-1}^\text{input} = \bm{x}_{t-1}^\text{input} + \frac{1}{n} \sum_{i\in\mathcal{M}} \bm{x}_{t-1,i}^\text{gen} - \frac{1}{n} \sum_{i\in\mathcal{M}} \bm{x}_{t-1,i}^\text{input}$ before creating the new combined representation $\bm{z}_{t-1} = [\tilde{\bm{z}}_{t-1}^\text{input}, \bm{z}_{t-1, i\notin\mathcal{M}}^\text{gen}]$ with $\tilde{\bm{z}}_{t-1}^\text{input} = [\tilde{\bm{x}}_{t-1}^\text{input}, \bm{h}_{t-1}^\text{input}]$ and $n=|\mathcal{M}|$.

\paragraph{Resampling}

\citet{trippe2022diffusion} show that this simple \textit{replacement method} inevitably introduces approximation error that can lead to inconsistent inpainted regions. In our experiments, we observe that the inpainting solution sometimes generates disconnected molecules that are not properly positioned in the target pocket (see Figure~\ref{fig:com_rmsd_resampling} for an example). \citet{trippe2022diffusion} propose to address this limitation with a particle filtering scheme that upweights more consistent samples in each denoising step. We, however, choose to adopt the conceptually simpler idea of \textit{resampling}~\cite{lugmayr2022repaint}, where each latent representation is repeatedly diffused back and forth before advancing to the next time step as demonstrated in Algorithm~\ref{alg:resampling}. This enables the model to harmonize its prediction for the generated part and the noisy sample from the fixed part, which does not include any information about the generated part. We choose $r=10$ resamplings per denoising step for our experiments with DiffSBDD-joint based on empirical results discussed in Appendix~\ref{supp:resampling}.

\begin{algorithm}[t]
    \caption{Sampling with the replacement method and resampling iterations. $r$ denotes the number of resampling steps and $\mathcal{M}$ is a set of indices of all atoms we want to fix.  Note that samples from the generative process $p_\theta(\bm{z}_{t-1}|\bm{z}_t)$ are assumed to be CoM-free.}\label{alg:resampling}
    \begin{algorithmic}
        \Require $r$, $\mathcal{M}$
        \State $\bm{z}_T \sim \mathcal{N}(\bm{0}, \bm{I})$

        \For{$t=T,...,1$}
            \For{$k=1,...,r$}

                \State $\bm{z}_{t-1}^\text{input} \sim q(\bm{z}_{t-1}|\bm{z}_\text{data})$ 
                \Comment{Sample known context}
                
                \State $\bm{z}_{t-1}^\text{gen} \sim p_\theta(\bm{z}_{t-1}|\bm{z}_t)$
                \Comment{Sample generated part}

                \State $\tilde{\bm{x}}_{t-1}^\text{input} = \bm{x}_{t-1}^\text{input} + \frac{1}{n} \sum_{i\in\mathcal{M}} \bm{x}_{t-1,i}^\text{gen} - \frac{1}{n} \sum_{i\in\mathcal{M}} \bm{x}_{t-1,i}^\text{input}$ 
                \Comment{Adjust center of mass}
                
                \State $\bm{z}_{t-1} = [\tilde{\bm{z}}_{t-1}^\text{input}, \bm{z}_{t-1, i\notin\mathcal{M}}^\text{gen}]$
                \Comment{Combine}
                
                \If{$k < r$}
                    \State $\bm{z}_t \sim q(\bm{z}_t|\bm{z}_{t-1})$
                    \Comment{Apply noise and repeat}
                \EndIf
                
            \EndFor
        \EndFor
        
    \State \Return $\bm{z}_0$
    \end{algorithmic}
\end{algorithm}

\subsection{Implementation details}\label{sec:implementation_details}

\paragraph{Molecule size}

As part of a sample's overall likelihood, we compute the empirical joint distribution of ligand and pocket nodes $p(N_L, N_P)$ observed in the training set and smooth it with a Gaussian filter ($\sigma=1$). In the conditional generation scenario, we derive the distribution $p(N_L | N_P)$ and use it for likelihood computations.

For sampling, we can either fix molecule sizes manually or sample the number of ligand nodes from the same distribution given the number of nodes in the target pocket:
\begin{equation}
    N_L \sim p(N_L | N_P).
\end{equation}

For the experiments discussed in Section~\ref{sec:distribution_learning}, we increase the mean size of sampled molecules by 5 (CrossDocked) and 10 (Binding MOAD) atoms, respectively, to approximately match the sizes of molecules found in the test set. This modification makes the reported Vina scores more comparable as the \textit{in silico} docking score is highly correlated with the molecular size, which is demonstrated in Figure~\ref{fig:qvina_score_correlation}. 
Average molecule sizes after applying the correction are shown in Table~\ref{tab:molecule_sizes} together with corresponding values for generated molecules from other methods.

\paragraph{Preprocessing}

All molecules are expressed as graphs. The full atom model uses the same one hot encoding of atom types for ligand and protein nodes. For the $C_\alpha$ only model the node features of the protein are set as the one hot encoding of the amino acid type instead. We refrain from adding a categorical feature for distinguishing between protein and ligand atoms and use two separate MLPs for embedding the node features instead (see Figure~\ref{fig:methods}F).

\paragraph{Noise schedule}
We use the pre-defined polynomial noise schedule introduced in \cite{hoogeboom2022equivariant}:
\begin{equation}
    \tilde{\alpha}_t = 1 - \Big(\frac{t}{T}\Big)^2, \quad t=0,...,T
\end{equation}
Following \cite{nichol2021improved, hoogeboom2022equivariant}, values of $\tilde{\alpha}_{t|s}^2=\big(\frac{\tilde{\alpha}_t}{\tilde{\alpha}_s}\big)^2$ are clipped between 0.001 and 1 for numerical stability near $t=T$, and $\tilde{\alpha}_t$ is recomputed as
\begin{equation}
    \tilde{\alpha}_t = \prod_{\tau=0}^t \tilde{\alpha}_{\tau|\tau - 1}.
\end{equation}
A tiny offset $\epsilon=10^{-5}$ is used to avoid numerical problems at $t=0$ defining the final noise schedule: 
\begin{equation}
    \alpha_t^2 = (1 - 2\epsilon) \cdot \tilde{\alpha}_t^2 + \epsilon.
\end{equation}

\paragraph{Feature scaling}
We scale the node type features $\bm{h}$ by a factor of 0.25 relative to the coordinates $\bm{x}$ which was empirically found to improve model performance in previous work~\cite{hoogeboom2022equivariant}. To train joint probability models in the all-atom scenario, it was necessary to scale down the coordinates (and corresponding distance cutoffs) by a factor of 0.2 instead in order to avoid introducing too many edges in the graph near the end of the diffusion process at $t=T$.

\paragraph{Hyperparameters}

Hyperparameters for all presented models are summarized in Table~\ref{tab:hyperparameters}.
Training takes about \SI{2.5}{\hour}/\SI{3.8}{\hour} (conditional/joint) per 100 epochs on a single NVIDIA V100 for Binding MOAD in the $C_\alpha$ scenario and \SI{11.5}{\hour}/\SI{14.7}{\hour} per 100 epochs with full atom pocket representation on two V100 GPUs. 
For CrossDocked, 100 training epochs take approximately \SI{6}{\hour}/\SI{8}{\hour} in the $C_\alpha$ case and \SI{48}{\hour}/\SI{60}{\hour} per 100 epochs on a single NVIDIA A100 GPU with all atom pocket representation.

\begin{table*}[h!]
    \caption{DiffSBDD hyperparameters.}
    \label{tab:hyperparameters}
    \centering
    \begin{adjustbox}{width=1\textwidth}
    \begin{tabular}{lcccccccc}
        \toprule
        & \multicolumn{4}{c}{CrossDocked} & \multicolumn{4}{c}{Binding MOAD} \\
        \cmidrule(r){2-9}
        & Cond & Joint & Cond ($C_\alpha$) & Joint ($C_\alpha$) & Cond & Joint & Cond ($C_\alpha$) & Joint ($C_\alpha$) \\
        \midrule
        No. layers & 5 & 5 & 6 & 6 & 6 & 6 & 5 & 5 \\
        Joint embedding dim. & 32 & 32 & 128 & 128 & 128 & 128 & 32 & 32 \\
        Hidden dim. & 128 & 128 & 256 & 256 & 192 & 192 & 128 & 128 \\
        Learning rate & $10^{-3}$ & $10^{-3}$ & $10^{-3}$ & $10^{-3}$ & $5\cdot 10^{-4}$ & $5\cdot 10^{-4}$ & $5\cdot 10^{-4}$ & $5\cdot 10^{-4}$ \\
        Weight decay & $10^{-12}$ & $10^{-12}$ & $10^{-12}$ & $10^{-12}$ & $10^{-12}$ & $10^{-12}$ & $10^{-12}$ & $10^{-12}$ \\
        Diffusion steps & 500 & 500 & 500 & 500 & 500 & 500 & 500 & 500 \\
        Edges (ligand-ligand) &  fully connected & fully connected & fully connected & fully connected & fully connected & fully connected & fully connected & fully connected \\
        Edges (ligand-pocket) & $< \SI{5}{\angstrom}$ & $< \SI{5}{\angstrom}$ & $< \SI{5}{\angstrom}$ & $< \SI{5}{\angstrom}$ & $< \SI{7}{\angstrom}$ & $< \SI{7}{\angstrom}$ & $< \SI{8}{\angstrom}$ & $< \SI{8}{\angstrom}$ \\
        Edges (pocket-pocket) & $< \SI{5}{\angstrom}$ & $< \SI{5}{\angstrom}$ & $< \SI{5}{\angstrom}$ & $< \SI{5}{\angstrom}$ & $< \SI{4}{\angstrom}$ & $< \SI{4}{\angstrom}$ & $< \SI{8}{\angstrom}$ & $< \SI{8}{\angstrom}$ \\
        Epochs & 1000 & 1000 & 1000 & 1000 & 1000 & 1000 & 1000 & 1000 \\
        \bottomrule
    \end{tabular}
    \end{adjustbox}
\end{table*}

\paragraph{Postprocessing}

For postprocessing of generated molecules, we use a similar procedure as in \cite{luo20213d}. Given a list of atom types and coordinates, bonds are first added using OpenBabel~\cite{o2011open}. 
We then use RDKit to sanitise molecules and filter for the largest molecular fragment.

\subsection{Experimental Set-up}
\label{sec:method_experiments}

\paragraph{Datasets} 
We use the CrossDocked dataset \cite{francoeur2020three} with 100,000 high-quality protein-ligand pairs for training and 100 proteins for testing, following previous works~\cite{luo20213d,peng2022pocket2mol}. The data split was done by 30\% sequence identity using MMseqs2~\cite{Steinegger2017}.

We also evaluate our method on a curated dataset of experimentally determined complexed protein-ligand structures from Binding MOAD~\cite{hu2005binding}.
We keep pockets with valid\footnote{as defined in \url{http://www.bindingmoad.org/}} and moderately `drug-like' ligands with QED score $>0.3$. We further discard small molecules that contain atom types $\notin \{C, N, O, S, B, Br, Cl, P, I, F\}$ as well as binding pockets with non-standard amino acids. We define binding pockets as the set of residues that have any atom within \SI{8}{\angstrom} of any ligand atom.
Ligand redundancy is reduced by randomly sampling at most 50 molecules with the same chemical component identifier (3-letter-code). After removing corrupted entries that could not be processed, \num{40344} training pairs and 130 testing pairs remain. A validation set of size 246 is used to monitor estimated log-likelihoods during training. The split is made to ensure different sets do not contain proteins from the same Enzyme Commission Number (EC Number) main class.

Since the ResGen baseline model did not successfully generate samples for 21 and 5 targets from the CrossDocked and Binding MOAD test sets, respectively, our analyses are only performed for samples from the remaining pockets.

\paragraph{Baselines} 

We select two recently published autoregressive deep-learning methods for structure-based drug design. Pocket2Mol~\cite{peng2022pocket2mol} and ResGen~\citep{zhang2023resgen} are sequential schemes relying on graph representations of the protein pocket and previously placed atoms to predict probabilities based on which new atoms are added.
They are currently state-of-the-art among this class of models.
Similar methods (e.g. 3D-SBDD~\cite{luo20213d} or GraphBP~\cite{liu2022generating}) consistently under-performed these baselines in our initial experiments and are therefore not included in the final comparison.
For Pocket2Mol, we re-evaluate already generated ligands on the CrossDocked dataset kindly provided by the authors. All other results were produced using the official implementations available online\footnote{\url{https://github.com/pengxingang/Pocket2Mol}, \url{https://github.com/HaotianZhangAI4Science/ResGen}} with default sampling parameters. Note that, unlike DiffSBDD, we therefore sample for the Binding MOAD test set with Pocket2Mol and ResGen models that have been trained on CrossDocked. Since these two sets overlap (30 test PDBs from Binding MOAD are found in the CrossDocked training set), there is potential data leakage. In practice, however, we do not observe significantly different results when these targets are excluded from the analysis.
We also attempted to train Pocket2Mol on Binding MOAD, but did not manage to robustly train the model on this dataset due to instability during training. 
While we aimed to sample 100 ligands per pocket for the results in Section~\ref{sec:distribution_learning}, the exact number of available molecules varies slightly due to the characteristics of the different methods (see Table~\ref{tab:num_molecules}).

\paragraph{Evaluation metrics} 

We employ widely-used metrics to assess the quality of our generated molecules \cite{peng2022pocket2mol, li2021structure}: (1)~\textbf{Vina Score} is an empirical estimation of binding affinity between small molecules and their target pocket; (2)~\textbf{QED} is a simple quantitative estimation of drug-likeness combining several desirable molecular properties~\citep{bickerton2012quantifying}; (3)~\textbf{SA} estimates synthetic accessibility, i.e. the difficulty of synthesis~\citep{ertl2009estimation}; (4)~\textbf{Lipinski} measures how many rules in the Lipinski rule of five~\cite{lipinski2012experimental} are satisfied (in addition to the original four rules we require 10 or fewer rotatable bonds); (5)~\textbf{Diversity} is computed as the average pairwise dissimilarity (1 - \textit{Tanimoto similarity}) between molecular fingerprints of all generated molecules for each pocket; (6)~\textbf{Inference Time} is the average sampling time per target. Roughly 100-130 molecules were sampled per target but the exact number varies between targets for Pocket2Mol and ResGen.
Chemical properties are calculated with RDKit~\cite{landrum2016rdkit}.
Docking scores are obtained after local minimization with an empirical force field using the GNINA implementation~\citep{mcnutt2021gnina} or, if specified, after redocking with QuickVina2~\cite{alhossary2015fast}.

\backmatter

\bmhead{Supplementary Information}
Description of method details and additional experimental results are included in the Supplementary Information.

\bmhead{Code Availability} Our source codes are publicly available at \url{https://github.com/arneschneuing/DiffSBDD}. Model weights can be downloaded from Zenodo: \url{https://zenodo.org/records/8183747}.

\bmhead{Data Availability} 
The subset of the CrossDocked dataset used in this study was curated in a previous work and is available online: \url{https://github.com/pengxingang/Pocket2Mol/tree/main/data}. The raw BindingMOAD data can be downloaded from \url{http://www.bindingmoad.org/}. We provide further instructions on how to process these data in our code repository: \url{https://github.com/arneschneuing/DiffSBDD}.
Sampled molecules are available on Zenodo: \url{https://zenodo.org/records/8239058}.

\bmhead{Acknowledgments}

We thank Xingang Peng and Shitong Luo for providing us generated molecules from Pocket2Mol.
We thank Hannes St\"{a}rk and Joshua Southern for valuable feedback and insightful discussions.
This work was supported by the European Research Council (starting grant no. 716058), the Swiss National Science Foundation (grant no. 310030\_188744), and Microsoft Research AI4Science.
Charles Harris is supported by the Cambridge Centre for AI in Medicine Studentship which is in turn funded by AstraZeneca and GSK. Michael Bronstein is supported in part by ERC Consolidator grant no. 724228 (LEMAN).

\bibliography{bibliography}%

\newpage
\begin{appendices}

\section{Note on Variational Lower Bound}
\label{supp:ELBO}
To maximise the likelihood of our training data, we aim at optimising the variational lower bound (VLB)~\cite{kingma2021variational,hoogeboom2022equivariant}
\begin{equation}
    -\log p(\bm{z}_\text{data}) \leq \underbrace{D_\text{KL}\big(q(\bm{z}_T|\bm{z}_\text{data})||p(\bm{z}_T)\big)}_\text{prior loss $\mathcal{L}_\text{prior}$} \underbrace{-\mathbb{E}_{q(\bm{z}_0|\bm{z}_\text{data})}\big[\log p(\bm{z}_\text{data}|\bm{z}_0)\big]}_\text{reconstruction loss $\mathcal{L}_0$} +  \underbrace{\sum_{t=1}^T\mathcal{L}_t}_\text{diffusion loss}
\end{equation}
with
\begin{align}
    \mathcal{L}_t &= D_\text{KL}\big(q(\bm{z}_{t-1}|\bm{z}_\text{data},\bm{z}_t)||p_\theta(\bm{z}_{t-1}|\hat{\bm{z}}_\text{data},\bm{z}_t)\big) \\
    &= \mathbb{E}_{\bm{\epsilon}\sim \mathcal{N}(\bm{0}, \bm{I})} \Big[\frac{1}{2}\Big(\frac{\text{SNR}(t-1)}{\text{SNR}(t)}-1\Big)||\bm{\epsilon}-\hat{\bm{\epsilon}}_\theta||^2\Big]
\end{align}
during training.
The prior loss should always be close to zero and can be computed exactly in closed form while the reconstruction loss must be estimated as described in \citet{hoogeboom2022equivariant}. In practice, however, we simply minimise the mean squared error $\mathcal{L}_\text{train} = \frac{1}{2}||\bm{\epsilon}-\hat{\bm{\epsilon}}||^2$ while randomly sampling time steps $t\sim \mathcal{U}(0,\dots,T)$, which is equivalent up to a multiplicative factor.

\section{Note on Equivariance of the Conditional Model}
The 3D-conditional model can achieve equivariance without the usual ``subspace-trick''. The coordinates of pocket nodes provide a reference frame for all samples that can be used to translate them to a unique location (e.g. such that the pocket is centered at the origin: $\sum_i \bm{x}^{(P)}_i = \bm{0}$). By doing this for all training data, translation equivariance becomes irrelevant and the CoM-free subspace approach obsolete. 
To evaluate the likelihood of translated samples at inference time, we can first subtract the pocket's center of mass from the whole system and compute the likelihood after this mapping. Similarly, for sampling molecules we can first generate a ligand in a CoM-free version of the pocket and move the whole system back to the original location of the pocket nodes to restore translation equivariance. 
As long as the mean of our Gaussian noise distribution $p(\bm{z}_t|\bm{z}^{(P)}_\text{data})=\mathcal{N}(\bm{\mu}(\bm{z}^{(P)}_\text{data}),\sigma^2\bm{I})$ depends equivariantly on the pocket node coordinates $\bm{x}^{(P)}$, $O(3)$-equivariance is satisfied as well (Appendix~\ref{sec:proofs}).
Since this change did not seem to affect the performance of the conditional model in our experiments, we decided to keep sampling in the linear subspace to ensure that the implementation is as similar as possible to the joint model, for which the subspace approach is necessary.

\section{Resampling}\label{supp:resampling}

Here, we briefly recapitulate the resampling algorithm introduced in Ref.~\cite{lugmayr2022repaint}. The key intuition is that inpainting with the replacement method combines a generated part with an independently sampled latent representation of the known part. Even though the neural network tries to reconcile these two components in every step of the denoising trajectory, it cannot succeed because the same issue reoccurs in the following step. \citet{lugmayr2022repaint} thus propose to apply the neural network several times before proceeding to the next noise level, allowing the DDPM to preserve more conditional information and move the sample closer to the data distribution again.

\paragraph{Number of resampling steps}

To empirically study the effect of the number of resampling iterations applied, we generated ligands for all test pockets with $r=1$, $r=5$, and $r=10$ resampling steps, respectively. Because the resampling strategy slows down sampling approximately by a factor of $r$, we used the striding technique proposed by \citet{nichol2021improved} and reduced the number of denoising steps proportionally to $r$. \citet{nichol2021improved} showed that this approach reduces the number of sampling steps significantly without sacrificing sample quality. In our case, it allows us to retain sampling speed while increasing the number of resampling steps.

\begin{figure}
    \centering
    \includegraphics[width=0.95\textwidth]{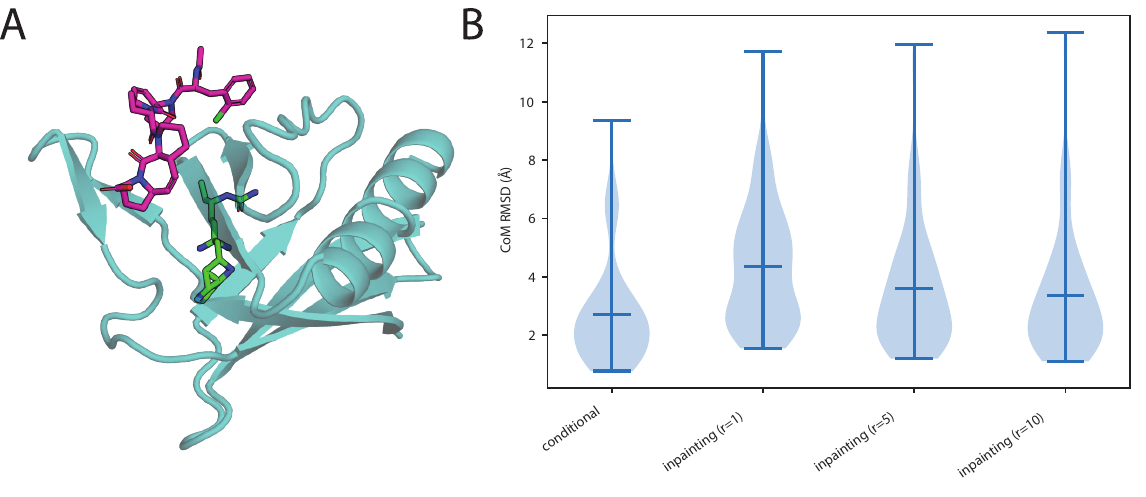}
    \caption{(A) Example of a generated molecule (green) without additional resampling steps and the reference molecule (magenta) from the target PDB 5ncf. The generated molecule is not placed in the target pocket but in the protein core. (B) RMSD between reference molecules' center of mass and generated molecules' center of mass for the conditional model and inpaining model with varying numbers of resampling steps $r$. The pocket representation is $C_\alpha$ in all cases.}
    \label{fig:com_rmsd_resampling}
\end{figure}

To gauge the effect of resampling for molecule generation we show the distribution of RMSD values between the center of mass of reference molecules and generated molecules in Figure~\ref{fig:com_rmsd_resampling}. The unmodified replacement method ($r=1$) produces molecules that are clearly farther away from the presumed pocket center than the conditional model. Increasing $r$ moves the mean distance closer to the average displacement of molecules from the conditional method. This effect seems to saturate at $r=10$ which is in line with the results obtained for images~\cite{lugmayr2022repaint}.

Table~\ref{tab:molecule_properties_bindingmoad_resampling_steps} shows that neither the additional resampling steps nor the shortened denoising trajectory degrade the performance on the reported molecular metrics. The average docking scores even improve slightly which might reflect better positioning of generated ligands in the pockets prior to docking. The same model trained with $T=500$ diffusion steps was used in all three cases.

\begin{table*}[t]
    \centering
    \caption{Evaluation of generated molecules for target pockets from the CrossDocked (C.D.) and Binding MOAD (B.M.) test sets with the inpainting approach and $C_\alpha$ pocket representation for varying numbers of resampling steps $r$ and denoising steps $T$. Here, the SA scores were mapped to the unit interval using $\text{SA}_\text{norm} = (10 - \text{SA}) / 9$.}
    \label{tab:molecule_properties_bindingmoad_resampling_steps}
    \begin{adjustbox}{width=1\textwidth}
    \begin{tabular}{ccccccccc}
    \toprule
    & $r$ & $T$ & Vina Score (kcal/mol, $\downarrow$) & QED ($\uparrow$) & $\text{SA}_\text{norm}$ ($\uparrow$) & Lipinski ($\uparrow$) & Diversity ($\uparrow$) & Time (s, $\downarrow$) \\
    \midrule
    \multirow{3}{*}{\rotatebox[origin=c]{90}{C.D.}}
    & 1 & 500 & $-5.830 \pm 2.47$ & $0.403 \pm 0.18$ & $0.552 \pm 0.13$ & $4.620 \pm 0.81$ & $0.808 \pm 0.06$ & $97.434 \pm 39.79$\\
    & 5 & 100 & $-6.872 \pm 2.43$ & $0.444 \pm 0.19$ & $0.551 \pm 0.12$ & $4.654 \pm 0.72$ & $0.766 \pm 0.06$ & $96.205 \pm 39.22$ \\
    & 10 & 50 & $-7.177 \pm 3.28$ & $0.556 \pm 0.20$ & $0.729 \pm 0.12$ &  $4.742 \pm 0.59$ & $0.718 \pm 0.07$ & $94.481 \pm 38.86$\\

    \midrule
    \multirow{3}{*}{\rotatebox[origin=c]{90}{B.M.}}
    & 1 & 500 & $-5.810 \pm 2.00$ & $0.468 \pm 0.16$ & $0.627 \pm 0.14$ & $4.839 \pm 0.49$ & $0.851 \pm 0.04$ & $40.298 \pm 13.52$ \\ 
    & 5 & 100 & $-6.082 \pm 2.01$ & $0.537 \pm 0.16$ & $0.701 \pm 0.13$ & $4.924 \pm 0.31$ & $0.855 \pm 0.05$ & $45.074 \pm 21.14$ \\ 
    & 10 & 50 & $-6.192 \pm 2.24$ & $0.560 \pm 0.16$ & $0.737 \pm 0.13$ & $4.941 \pm 0.27$ & $0.859 \pm 0.05$ & $41.490 \pm 14.32$ \\
    \bottomrule
    \end{tabular}
    \end{adjustbox}
\end{table*}

\section{Proofs}\label{sec:proofs}

In the following proofs we do not consider categorical node features $\bm{h}$ as only the positions $\bm{x}$ are subject to equivariance constraints. Furthermore, we do not distinguish between the zeroth latent representation $\bm{x}_0$ and data domain representations $\bm{x}_\text{data}$ for ease of notation, and simply drop the subscripts.

\subsection{$O(3)$-equivariance of the prior probability}
\label{sec:o3-equivariance-prior}

The isotropic Gaussian prior $p(\bm{x}^{(L)}_T | \bm{x}^{(P)}) = \mathcal{N}(\bm{\mu}(\bm{x^{(P)}}), \sigma^2\bm{I})$ is equivariant to rotations and reflections represented by an orthogonal matrix $\bm{R}\in\mathbb{R}^{3 \times 3}$ as long as $\bm{\mu}(\bm{R}\bm{x}^{(P)}) = \bm{R}\bm{\mu}(\bm{x}^{(P)})$ because:
\begin{align*}
    p(\bm{R}\bm{x}^{(L)}_T | \bm{R}\bm{x}^{(P)}) &= \frac{1}{\sqrt{(2\pi)^{N_L}\sigma^2}} \exp\Big(-\frac{1}{2\sigma^2} || \bm{R}\bm{x}^{(L)}_T - \bm{\mu}(\bm{R}\bm{x}^{(P)}) ||^2\Big) \\
    &= \frac{1}{\sqrt{(2\pi)^{N_L}\sigma^2}} \exp\Big(-\frac{1}{2\sigma^2} || \bm{R}\bm{x}^{(L)}_T - \bm{R}\bm{\mu}(\bm{x}^{(P)}) ||^2\Big) \\
    &= \frac{1}{\sqrt{(2\pi)^{N_L}\sigma^2}} \exp\Big(-\frac{1}{2\sigma^2} || \bm{R}\big(\bm{x}^{(L)}_T - \bm{\mu}(\bm{x}^{(P)})\big) ||^2\Big) \\
    &= \frac{1}{\sqrt{(2\pi)^{N_L}\sigma^2}} \exp\Big(-\frac{1}{2\sigma^2} || \bm{x}^{(L)}_T - \bm{\mu}(\bm{x}^{(P)}) ||^2\Big) \\
    &= p(\bm{x}^{(L)}_T | \bm{x}^{(P)}).
\end{align*}

Here we used $||\bm{R}\bm{x}||_2 = ||\bm{x}||_2$ for orthogonal $\bm{R}$.

\subsection{$O(3)$-equivariance of the transition probabilities}
The denoising transition probabilities from time step $t$ to $s < t$ are defined as isotropic normal distributions:
\begin{equation}
    p_\theta(\bm{x}^{(L)}_{t-1}|\bm{x}^{(L)}_t, \hat{\bm{x}}^{(L)}, \bm{x}^{(P)}) = \mathcal{N}(\bm{x}^{(L)}_{t-1} | \bm{\mu}_{t\rightarrow s}(\bm{x}^{(L)}_t, \hat{\bm{x}}^{(L)}, \bm{x}^{(P)}), \sigma_{t\rightarrow s}^2 \bm{I}).
\end{equation}
Therefore, $p_\theta(\bm{x}^{(L)}_{t-1}|\bm{x}^{(L)}_t, \hat{\bm{x}}^{(L)}, \bm{x}^{(P)})$ is $O(3)$-equivariant by a similar argument to Section~\ref{sec:o3-equivariance-prior} if $\bm{\mu}_{t\rightarrow s}$ is computed equivariantly from the three-dimensional context.

Recalling the definition of $\bm{\mu}_{t\rightarrow s} = \frac{\alpha_{t|s}\sigma_s^2}{\sigma_t^2}\bm{x}^{(L)}_t + \frac{\alpha_s \sigma_{t|s}^2}{\sigma_t^2}\hat{\bm{x}}^{(L)}$, we can prove its equivariance as follows:
\begin{align*}
    \bm{\mu}_{t\rightarrow s}(\bm{R}\bm{x}^{(L)}_t, \bm{R}\bm{x}^{(P)}) 
    &= \frac{\alpha_{t|s}\sigma_s^2}{\sigma_t^2}\bm{R}\bm{x}^{(L)}_t + \frac{\alpha_s \sigma_{t|s}^2}{\sigma_t^2}\hat{\bm{x}}^{(L)}(\bm{R}\bm{x}_t^{(L)}, \bm{R}\bm{x}^{(P)}) \\
    &= \frac{\alpha_{t|s}\sigma_s^2}{\sigma_t^2}\bm{R}\bm{x}^{(L)}_t + \frac{\alpha_s \sigma_{t|s}^2}{\sigma_t^2}\bm{R}\hat{\bm{x}}^{(L)}(\bm{x}_t^{(L)}, \bm{x}^{(P)}) \quad \text{(equivariance of $\hat{\bm{x}}^{(L)}$)} \\
    &= \bm{R} \Big( \frac{\alpha_{t|s}\sigma_s^2}{\sigma_t^2}\bm{x}^{(L)}_t + \frac{\alpha_s \sigma_{t|s}^2}{\sigma_t^2}\hat{\bm{x}}^{(L)}(\bm{x}_t^{(L)}, \bm{x}^{(P)}) \Big) \\
    &= \bm{R} \bm{\mu}_{t\rightarrow s}(\bm{x}^{(L)}_t, \bm{x}^{(P)}),
\end{align*}

where $\hat{\bm{x}}^{(L)}$ defined as $\hat{\bm{x}}^{(L)}= \frac{1}{\alpha_t}\bm{x}_t^{(L)}-\frac{\sigma_t}{\alpha_t}\hat{\bm{\epsilon}}$ is equivariant because:
\begin{align*}
    \hat{\bm{x}}^{(L)}(\bm{R}\bm{x}_t^{(L)}, \bm{R}\bm{x}^{(P)}) 
    &= \frac{1}{\alpha_t}\bm{R}\bm{x}_t^{(L)}-\frac{\sigma_t}{\alpha_t}\hat{\bm{\epsilon}}(\bm{R}\bm{x}_t^{(L)}, \bm{R}\bm{x}^{(P)}, t) \\
    &= \frac{1}{\alpha_t}\bm{R}\bm{x}_t^{(L)}-\frac{\sigma_t}{\alpha_t}\bm{R}\hat{\bm{\epsilon}}(\bm{x}_t^{(L)}, \bm{x}^{(P)}, t) \quad \text{($\hat{\bm{\epsilon}}$ predicted by equivariant neural network)} \\
    &= \bm{R} \Big(\frac{1}{\alpha_t}\bm{x}_t^{(L)}-\frac{\sigma_t}{\alpha_t}\hat{\bm{\epsilon}}(\bm{x}_t^{(L)}, \bm{x}^{(P)}, t) \Big) \\
    &= \bm{R}\hat{\bm{x}}^{(L)}(\bm{x}_t^{(L)}, \bm{x}^{(P)}).
\end{align*}

\subsection{$O(3)$-equivariance of the learned likelihood}

Let $\bm{R}\in\mathbb{R}^{3 \times 3}$ be an orthogonal matrix representing an element $g$ from the general orthogonal group $O(3)$.
We obtain the marginal probability density of the Markovian denoising process as follows
\begin{align*}
    p_\theta(\bm{x}^{(L)}_0|\bm{x}^{(P)}) &= \int p(\bm{x}^{(L)}_T|\bm{x}^{(P)}) p_\theta(\bm{x}^{(L)}_{0:T-1}|\bm{x}^{(L)}_T, \bm{x}^{(P)}) \text{d}\bm{x}_{1:T} \\ 
    &= \int p(\bm{x}^{(L)}_T|\bm{x}^{(P)}) \prod_{t=1}^T p_\theta(\bm{x}^{(L)}_{t-1}|\bm{x}^{(L)}_t, \bm{x}^{(P)}) \text{d}\bm{x}_{1:T}
\end{align*}

and the sample's likelihood is $O(3)$-equivariant:
\begin{align*}
    p_\theta(\bm{R}\bm{x}^{(L)}_0|\bm{R}\bm{x}^{(P)}) &= \int p(\bm{R}\bm{x}^{(L)}_T|\bm{R}\bm{x}^{(P)}) \prod_{t=1}^T p_\theta(\bm{R}\bm{x}^{(L)}_{t-1}|\bm{R}\bm{x}^{(L)}_t, \bm{R}\bm{x}^{(P)}) \text{d}\bm{x}_{1:T} \\
    &= \int p(\bm{x}^{(L)}_T|\bm{x}^{(P)}) \prod_{t=1}^T p_\theta(\bm{R}\bm{x}^{(L)}_{t-1}|\bm{R}\bm{x}^{(L)}_t, \bm{R}\bm{x}^{(P)}) \text{d}\bm{x}_{1:T} \quad \text{(equivariant prior)} \\
    &= \int p(\bm{x}^{(L)}_T|\bm{x}^{(P)}) \prod_{t=1}^T p_\theta(\bm{x}^{(L)}_{t-1}|\bm{x}^{(L)}_t, \bm{x}^{(P)}) \text{d}\bm{x}_{1:T} \quad \text{(equivariant transition probabilities)} \\
    &= p_\theta(\bm{x}^{(L)}_0|\bm{x}^{(P)})
\end{align*}

\section{$SE(3)$-equivariant Graph Neural Network}
\label{sec:SEGNN}
Chiral molecules cannot be superimposed by any combination of rotations and translations. Instead they are mirrored along a stereocenter, axis, or plane.
As chirality can significantly alter a molecule's chemical properties, we use a variant of the $E(3)$-equivariant graph neural networks~\cite{satorras2021n} presented in Equations \eqref{egnn_message}-\eqref{egnn_coordinate} that is sensitive to reflections and hence $SE(3)$-equivariant. We change the coordinate update equation, Equ.~\eqref{egnn_coordinate}, of standard EGNNs in the following way
\begin{equation}
    \bm{x}_{i}^{l+1} = \bm{x}_{i}^{l} + \sum_{j\neq i} \frac{\bm{x}_{i}^{l} - \bm{x}_{j}^{l}}{d_{ij} + 1}  \phi_x^d(\bm{h}^l_i, \bm{h}^l_j, d_{ij}^2, a_{ij}) + \frac{(\bm{x}_{i}^{l} - \bar{\bm{x}}^l) \times (\bm{x}_{j}^{l} - \bar{\bm{x}}^l)}{||(\bm{x}_{i}^{l} - \bar{\bm{x}}^l) \times (\bm{x}_{j}^{l} - \bar{\bm{x}}^l)|| + 1} \phi_x^\times(\bm{h}^l_i, \bm{h}^l_j, d_{ij}^2, a_{ij}), \label{equ:segnn_coordinate}
\end{equation}
where $\bar{\bm{x}}^l$ denotes the center of mass of all nodes at layer $l$.
This modification makes the EGNN layer sensitive to reflections while staying close to the original formalism. Since the resulting graph neural networks are only equivariant to the $SE(3)$ group, we will hereafter call them SE(3)GNNs for short.

\subsection{Discussion of Equivariance}
Here we study how the suggested change in the coordinate update equation breaks reflection symmetry while preserving equivariance to rotations. Messages and scalar feature updates (Equations~\eqref{egnn_message} and \eqref{egnn_node}) remain $E(3)$-invariant as in the original model and are therefore not considered in this section.
We analyze transformations composed of a translation by $\bm{t}\in\mathbb{R}^3$ and a rotation/reflection by an orthogonal matrix $\bm{R}\in\mathbb{R}^{3\times 3}$ with $\bm{R}^T\bm{R}=\bm{I}$. 
The output at layer $l+1$ given the transformed input $\bm{R}\bm{x}^l_i+\bm{t}$ at layer $l$ is calculated as:
{\small
\begin{align}
    \bm{R}\bm{x}^l_i+\bm{t} + \sum_{j\neq i} \frac{\bm{R}\bm{x}^l_i+\bm{t} - (\bm{R}\bm{x}^l_j+\bm{t})}{d_{ij} + 1}  \phi_x^d(\cdot) + \frac{(\bm{R}\bm{x}^l_i+\bm{t} - (\bm{R}\bar{\bm{x}}^l + \bm{t})) \times (\bm{R}\bm{x}^l_j + \bm{t} - (\bm{R}\bar{\bm{x}}^l + \bm{t}))}{Z^\times_{ij} + 1} \phi_x^\times(\cdot) \\
    = \bm{R}\bm{x}^l_i+\bm{t} + \sum_{j\neq i} \frac{\bm{R} (\bm{x}^l_i - \bm{x}^l_j)}{d_{ij} + 1}  \phi_x^d(\cdot) + \frac{(\bm{R}\bm{x}^l_i - \bm{R}\bar{\bm{x}}^l) \times (\bm{R}\bm{x}^l_j - \bm{R}\bar{\bm{x}}^l)}{Z^\times_{ij} + 1} \phi_x^\times(\cdot) \\
    = \bm{R}\bm{x}^l_i+\bm{t} + \sum_{j\neq i} \frac{\bm{R} (\bm{x}^l_i - \bm{x}^l_j)}{d_{ij} + 1}  \phi_x^d(\cdot) + \frac{\det(\bm{R})\bm{R}\big((\bm{x}^l_i - \bar{\bm{x}}^l) \times (\bm{x}^l_j - \bar{\bm{x}}^l)\big)}{Z^\times_{ij} + 1} \phi_x^\times(\cdot) \\
    = \bm{R}\bm{x}_i^{l+1} + \bm{t} + \big(\det(\bm{R})-1\big) \sum_{j\neq i} \frac{\bm{R}\big((\bm{x}^l_i - \bar{\bm{x}}^l) \times (\bm{x}^l_j - \bar{\bm{x}}^l)\big)}{Z^\times_{ij} + 1}.
\end{align}
}

This result shows that the output coordinates are only equivariantly transformed if $\bm{R}$ is orientation preserving, i.e. $\det(\bm{R})=1$. If $\bm{R}$ is a reflection ($\det(\bm{R})=-1$), coordinates will be updated with an additional summand that breaks the symmetry.

The learnable coefficients $\phi_x^d(\bm{h}^l_i, \bm{h}^l_j, d_{ij}^2, a_{ij})$ and $\phi_x^\times(\bm{h}^l_i, \bm{h}^l_j, d_{ij}^2, a_{ij})$ only depend on relative distances and are therefore $E(3)$-invariant. Their arguments are represented with the ``$\cdot$'' symbol for brevity. Likewise, the normalization factor $||(\bm{x}_{i}^{l} - \bar{\bm{x}}^l) \times (\bm{x}_{j}^{l} - \bar{\bm{x}}^l)||$ is abbreviated as $Z^\times_{ij}$. Already in the first line we used the fact that the mean transforms equivariantly. Furthermore, we use $\bm{R}\bm{a} \times \bm{R}\bm{b} = \det(\bm{R})\bm{R}(\bm{a} \times \bm{b})$ in the second step, which can be derived as follows:
\begin{align}
    \bm{x}^T(\bm{R}\bm{a}\times \bm{R}\bm{b}) &= \det(\underbrace{[\bm{x}, \bm{R}\bm{a}, \bm{R}\bm{b}]}_{\in \mathbb{R}^{3\times 3}}) \\
    &= \det(\bm{R}[\bm{R}^T\bm{x}, \bm{a}, \bm{b}]) \\
    &= \det(\bm{R})\det([\bm{R}^T\bm{x}, \bm{a}, \bm{b}]) \\
    &= \det(\bm{R}) \Big( \bm{x}^T\bm{R}(\bm{a}\times \bm{b}) \Big) \\
    &= \bm{x}^T\Big( \det(\bm{R})\bm{R}(\bm{a}\times \bm{b}) \Big)
\end{align}
The stated property of the cross product follows because this derivation is true for all $\bm{x}\in\mathbb{R}^{3}$.

\begin{wraptable}{r}{0.5\textwidth}
    \centering
    \vspace{-5mm}
    \caption{Accuracy on the R/S classification task. Results in the first section are taken from \cite{adams2021learning} and included for reference. 
    }
    \label{tab:rs_classification}
    \begin{tabular}{ lc }
        \toprule
        Model & R/S Accuracy (\%) \\
        \toprule
        ChIRo & $98.5$ \\ 
        SchNet & $54.4$ \\ 
        DimeNet++ & $65.7$ \\ 
        SphereNet & $98.2$ \\ 
        \midrule
        EGNN & $50.4$ \\
        SE(3)GNN & $83.4$ \\
        \bottomrule
    \end{tabular}
\end{wraptable}

\subsection{Empirical Results}
To show the effectiveness of this architecture on a simple toy example, we repeat the classification experiment by \citet{adams2021learning} who train neural networks to classify tetrahedral chiral centers as right-handed (\textit{rectus}, `R') or left-handed (\textit{sinister}, `S'). We closely follow their data split and experimental set-up and only replace the classifier with EGNN and SE(3)GNNs, respectively. The results in Table~\ref{tab:rs_classification} clearly demonstrate that the $SE(3)$-equivariant EGNN is capable of solving this task (without any hyperparameter optimization) whereas the $E(3)$-equivariant version does not do better than random guessing.

\section{Extended results}

\subsection{Sampling statistics}
Table~\ref{tab:num_molecules} summarizes the number of generated molecules available for the analyses in Section~\ref{sec:distribution_learning}.

\begin{table}[h]
    \centering
    \caption{Average number of molecules per test set pocket.}
    \label{tab:num_molecules}
    \begin{tabular}{lcc}
    \toprule
        Method & CrossDocked & Binding MOAD \\
        \midrule
        Test set & 1 & 1 \\
        \midrule
        Pocket2Mol~\cite{peng2022pocket2mol} & 98.01 & 132.14 \\
        ResGen~\cite{zhang2023resgen} & 114.41 & 109.83 \\
        \midrule
        DiffSBDD-cond & 100 & 100 \\
        DiffSBDD-joint & 100 & 100 \\
        \bottomrule
    \end{tabular}
\end{table}

\subsection{Distributions of Molecular Properties}\label{sec:score_distributions}

\begin{figure}
    \centering
    \includegraphics[width=\textwidth]{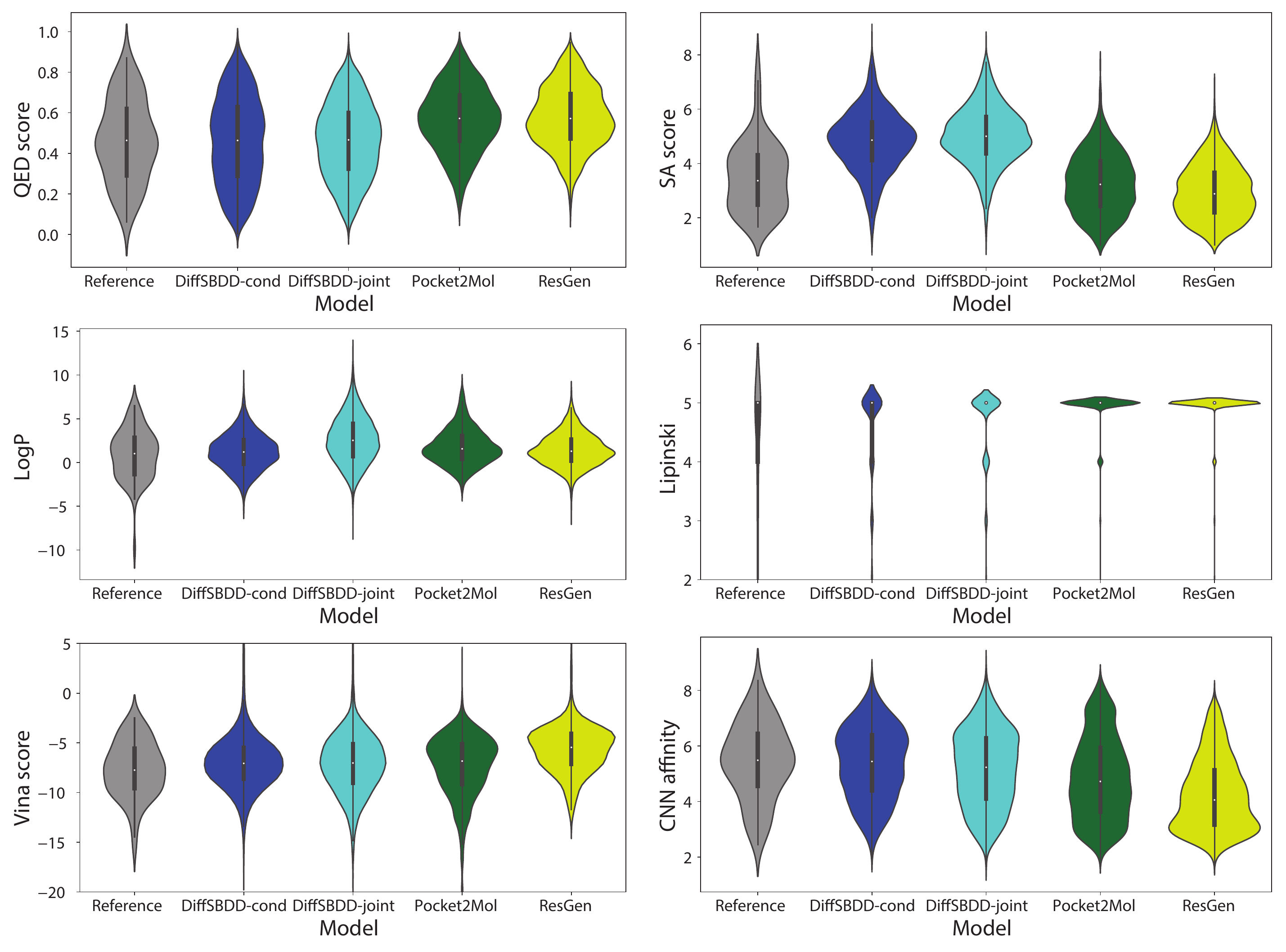}
    \caption{Distributions of computational scores for generated molecules and reference ligands from the CrossDocked test set.}
    \label{fig:distributions_crossdocked}
\end{figure}

\begin{figure}
    \centering
    \includegraphics[width=\textwidth]{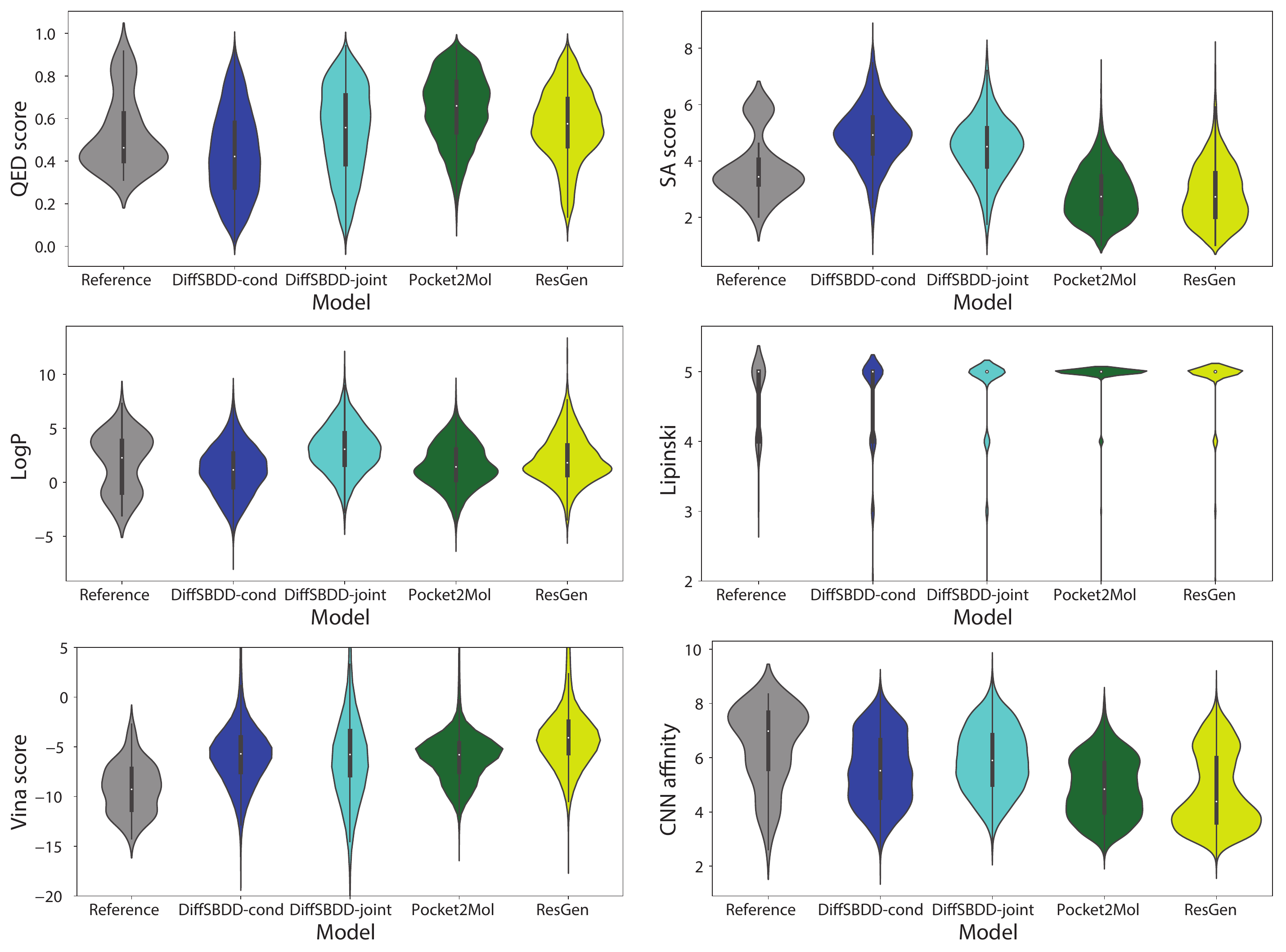}
    \caption{Distributions of computational scores for generated molecules and reference ligands from the Binding MOAD test set.}
    \label{fig:distributions_bindingmoad}
\end{figure}

The distributions used to compute the Wasserstein distances in Table~\ref{tab:table_distribution_learning} are visualized in Figures~\ref{fig:distributions_crossdocked} and \ref{fig:distributions_bindingmoad}.

\subsection{Dependence of Vina scores on molecule size}

\begin{figure}
    \centering
    \includegraphics[width=0.9\textwidth]{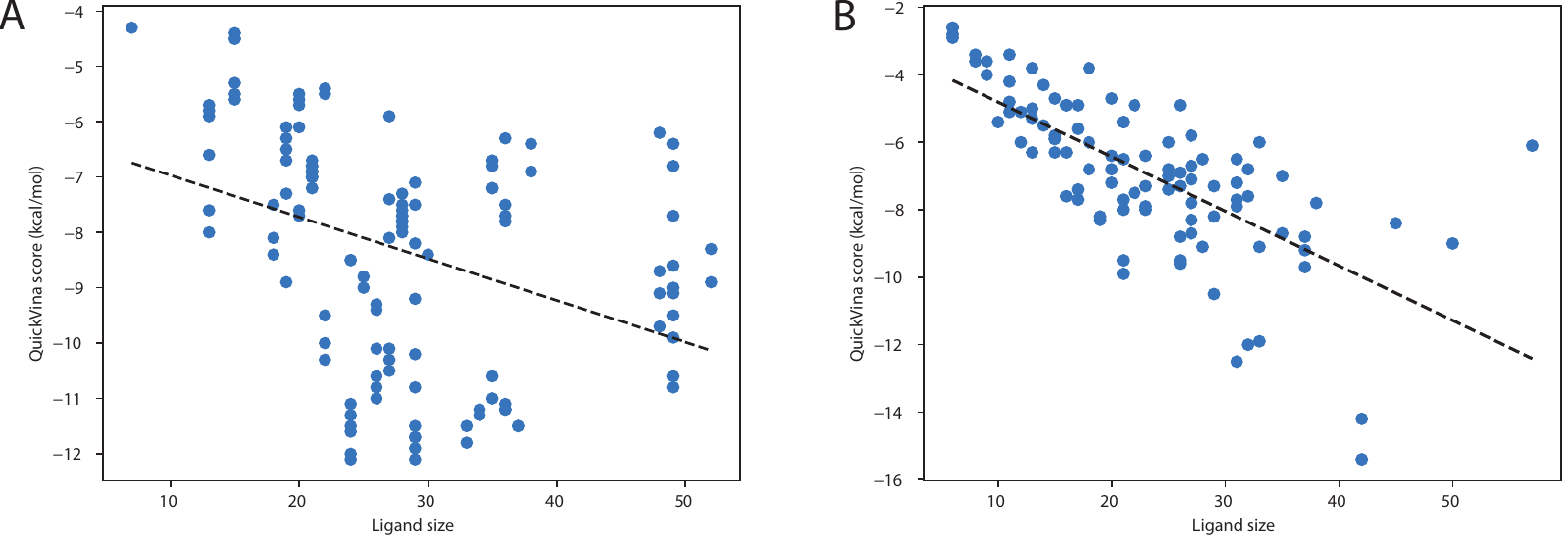}
    \caption{Correlation between ligand size and QuickVina score for reference molecules from the Binding MOAD (A) and CrossDocked (B) test sets.}
    \label{fig:qvina_score_correlation}
\end{figure}

Figure~\ref{fig:qvina_score_correlation} shows how strongly the empirical Vina score is correlated with the number of heavy atoms in the ligands. For this analysis, we used Vina scores computed by the QuickVina2~\cite{alhossary2015fast} software after re-docking. Since we want to match the distribution of scores of reference molecules as closely as possible with our generated molecules, we expect that their sizes should roughly match as well. However, our diffusion model operates on point clouds with fixed sizes, which we determine at the beginning of sampling as explained in Section~\ref{sec:implementation_details}. By biasing the procedure, we match the sizes of reference ligands more closely as shown in Table~\ref{tab:molecule_sizes}.

\begin{table}
    \centering
    \caption{Average number of heavy atoms of generated molecules.}
    \label{tab:molecule_sizes}
    \begin{tabular}{lcc}
    \toprule
        Method & CrossDocked & Binding MOAD \\
        \midrule
        Test set & $23.9$ & $28.0$ \\
        \midrule
        Pocket2Mol~\cite{peng2022pocket2mol} & $ 19.0$ & 16.8 \\
        ResGen~\cite{zhang2023resgen} & 16.2 & 18.4 \\
        \midrule
        DiffSBDD-cond & $24.9$ & $24.5$ \\
        DiffSBDD-joint & $24.5$ & $25.2$ \\
        \bottomrule
    \end{tabular}
\end{table}

\subsection{Additional Molecular Metrics}
In addition to the molecular properties discussed in Section~\ref{sec:method_experiments} we assess the models' ability to produce novel and valid molecules using four simple metrics: validity, connectivity, uniqueness, and novelty. \textbf{Validity} measures the proportion of generated molecules that pass basic tests by RDKit--mostly ensuring correct valencies. \textbf{Connectivity} is the proportion of valid molecules that do not contain any disconnected fragments. We convert every valid and connected molecule from a graph into a canonical SMILES string representation, count the number unique occurrences in the set of generated molecules and compare those to the training set SMILES to compute \textbf{uniqueness} and \textbf{novelty} respectively.

Table~\ref{tab:basic_metrics} shows that only a small fraction of all generated molecules is invalid and must be discarded for downstream processing.
A much larger percentage of molecules is fragmented but, since we can simply select and process the largest fragments in these cases, low connectivity does not necessarily affect the efficiency of the generative process.
Moreover, all models produce diverse sets of molecules unseen in the training set.

\begin{table*}[h!]
    \centering
    \caption{Basic molecular metrics for generated small molecules given a $C_\alpha$ and full atom representation of the protein pocket. }
    \label{tab:basic_metrics}
    \begin{tabular}{ lcccc } 
    \toprule
    \bf{Model} & \bf{Validity} & \bf{Connectivity} & \bf{Uniqueness} & \bf{Novelty} \\
    \midrule
    CrossDocked test set & 100\% & 100\% & 96\% & 96.88\% \\
    DiffSBDD-cond ($C_\alpha$) & 95.52\% & 79.52\% & 99.99\% & 99.97\%\\
    DiffSBDD-inpaint ($C_\alpha$)& 99.18\% & 98.25\% & 99.52\% & 99.97\%\\
    DiffSBDD-cond & 97.10\% & 78.27\% & 99.98\% & 99.99\%\\
    DiffSBDD-inpaint & 92.99\% & 67.52\% & 100\% & 100\%\\
    \midrule
    Binding MOAD test set  & 97.69\% & 100\% & 38.58\% & 77.55\% \\
    DiffSBDD-cond ($C_\alpha$) & 94.41\% & 77.38\% & 100\% & 100\% \\
    DiffSBDD-inpaint ($C_\alpha$) & 98.36\% & 91.60\% & 99.99\% & 99.98\% \\
    DiffSBDD-cond & 96.32\% & 63.37\% & 100\% & 100\% \\
    DiffSBDD-inpaint & 93.88\% & 75.60\% & 100\% & 100\% \\
    \bottomrule
    \end{tabular}
\end{table*}

\subsection{Results with a coarse-grained pocket representation}\label{sec:CA_models}

\begin{table*}[]
    \centering
    \caption{Evaluation of generated molecules for targets from the CrossDocked and Binding MOAD test sets. We compare all-atom and coarse-grained ($C_\alpha$) pocket representations. Note that the SA score has been normalized so that higher is better. Here, the SA scores were mapped to the unit interval using $\text{SA}_\text{norm} = (10 - \text{SA}) / 9$.}
    \label{tab:ca_models}
    \begin{adjustbox}{width=1\textwidth}
    \begin{tabular}{clccccccc}
    \toprule
    & & Vina (All) ($\downarrow$) & Vina (Top-10\%) ($\downarrow$) & QED ($\uparrow$) & $\text{SA}_\text{norm}$ ($\uparrow$) & Lipinski ($\uparrow$) & Diversity ($\uparrow$) & Time (s, $\downarrow$) \\
    \midrule
    \multirow{5}{*}{\rotatebox[origin=c]{90}{C.D.}}
    & Test set & $-6.871 \pm 2.32$ & --- & $0.476 \pm 0.20$ & $0.728 \pm 0.14$ & $4.340 \pm 1.14$ & --- & ---\\
    \cmidrule{2-9}
    & DiffSBDD-cond ($C_\alpha$) & $-6.770 \pm 2.73$ & $-8.796 \pm 1.75$ & $0.475 \pm 0.22$ & $0.612 \pm 0.12$ & $4.536 \pm 0.91$ & $0.725 \pm 0.06$ & \two{$49.651 \pm 17.34$} \\
    & DiffSBDD-inpaint ($C_\alpha$) & \three{$-7.177 \pm 3.28$} & \three{$-9.233 \pm 1.82$} & \two{$0.556 \pm 0.20$} & \two{$0.729 \pm 0.12$} &  $4.742 \pm 0.59$ & $0.718 \pm 0.07$ & \three{$94.481 \pm 38.86$} \\
    & DiffSBDD-cond & $-6.950 \pm 2.06$ & $-9.120 \pm 2.16$ & $0.469 \pm 0.21$ & $0.578 \pm 0.13$ & $4.562 \pm 0.89$ & $0.728 \pm 0.07$ & $135.866 \pm 51.66$\\
    & DiffSBDD-inpaint & \one{$-7.333 \pm 2.56$} & \one{$-9.927 \pm 2.59$} & $0.467\pm 0.18$ & $0.554 \pm 0.12$ & $4.702 \pm 0.64$ & \two{$0.758 \pm 0.05$} & $160.314 \pm 73.30$\\
    \midrule
    
    \multirow{6}{*}{\rotatebox[origin=c]{90}{B.M.}}
    & Test set & $-8.412 \pm 2.03$ & --- & $0.522 \pm 0.17$ & $0.692 \pm 0.12$ & $4.669 \pm 0.49$ & --- & --- \\
    \cmidrule{2-9}
    & DiffSBDD-cond ($C_\alpha$) & $-6.863 \pm 1.59$ & $-8.587 \pm 1.34$ & $0.480 \pm 0.20$ & $0.554 \pm 0.11$ & $4.662 \pm 0.68$ & \three{$0.714 \pm 0.05$} & \two{$36.285 \pm 8.13$} \\
    & DiffSBDD-inpaint ($C_\alpha$) & \three{$-6.926 \pm 3.39$} & \three{$-9.124 \pm 1.35$} & \one{$0.548 \pm 0.19$} & \two{$0.580 \pm 0.13$} & \three{$4.757 \pm 0.51$} & $0.709 \pm 0.05$ & \three{$58.305 \pm 17.35$} \\
    & DiffSBDD-cond & \two{$-7.171 \pm 1.89$} & \two{$-9.184 \pm 2.23$} & $0.436 \pm 0.20$ & \three{$0.568 \pm 0.12$} & $4.542 \pm 0.79$ & \three{$0.714 \pm 0.08$} & $336.061 \pm 85.02$ \\
    & DiffSBDD-inpaint & \one{$-7.309 \pm 4.03$} & \one{$-9.840 \pm 2.18$} & \two{$0.542 \pm 0.21$} & \one{$0.615 \pm 0.12$} & \two{$4.777 \pm 0.53$} & \two{$0.739 \pm 0.05$} & $369.873 \pm 124.54$ \\
    \bottomrule
    \end{tabular}
    \end{adjustbox}
\end{table*}

\begin{figure}
    \centering
    \includegraphics[width=\textwidth]{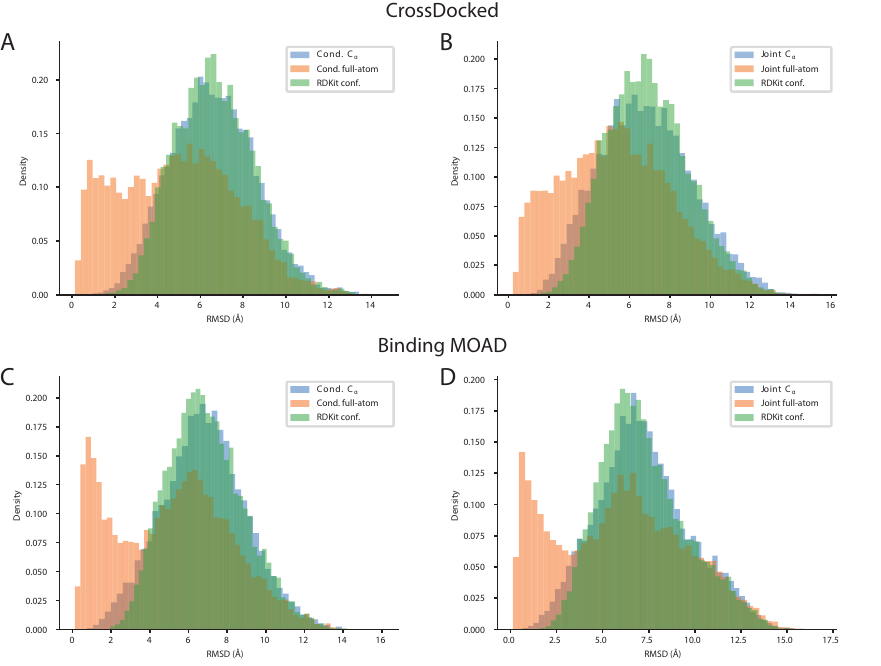}
    \caption{RMSD between generated and docked conformations for the CrossDocked (A, B) and Binding MOAD (C, D) datasets. Full-atom models are compared to $C_\alpha$ models as well as a baseline of random RDKit conformers of the molecules generated by the $C_\alpha$-model. (A, C) DiffSBDD-cond. (B) DiffSBDD-joint.}
    \label{fig:qvina_rmsd}
\end{figure}

Here we discuss the advantages and disadvantages of coarse-grained $C_\alpha$ protein representations as context for the generative model. Table~\ref{tab:ca_models} shows that full-atom models outperform their coarse-grained counterparts on the Vina metric, which is the only reported metric that captures interactions with the protein. Ligand-centric metrics do not seem to depend on the protein representation as could be expected. The main advantage of the $C_\alpha$ models is their significantly faster training and inference time, a fact we made use of during model development for fine-tuning and preliminary analyses.

To further demonstrate the limitations of the coarse-grained models, we 
compare the generated raw conformations to the best scoring QuickVina docking pose after re-docking and plot the distribution of resulting RMSD values in Figure~\ref{fig:qvina_rmsd}. As a baseline, the procedure is repeated for RDKit conformers of the same molecules with identical center of mass.
For a large percentage of molecules generated by the all-atom models, QuickVina agrees with the predicted bound conformations, leaving them almost unchanged (RMSD below \SI{2}{\angstrom}). This demonstrates successful conditioning on the geometry of the given protein pockets.
For the $C_\alpha$-only models results are less convincing. They produce poses that barely improve upon conformers lacking pocket-context. Likely, this is caused by atomic clashes with the proteins' side chains that QuickVina needs to resolve.

\subsection{Random generated molecules}

\begin{figure}
     \centering
     \includegraphics[width=\textwidth]{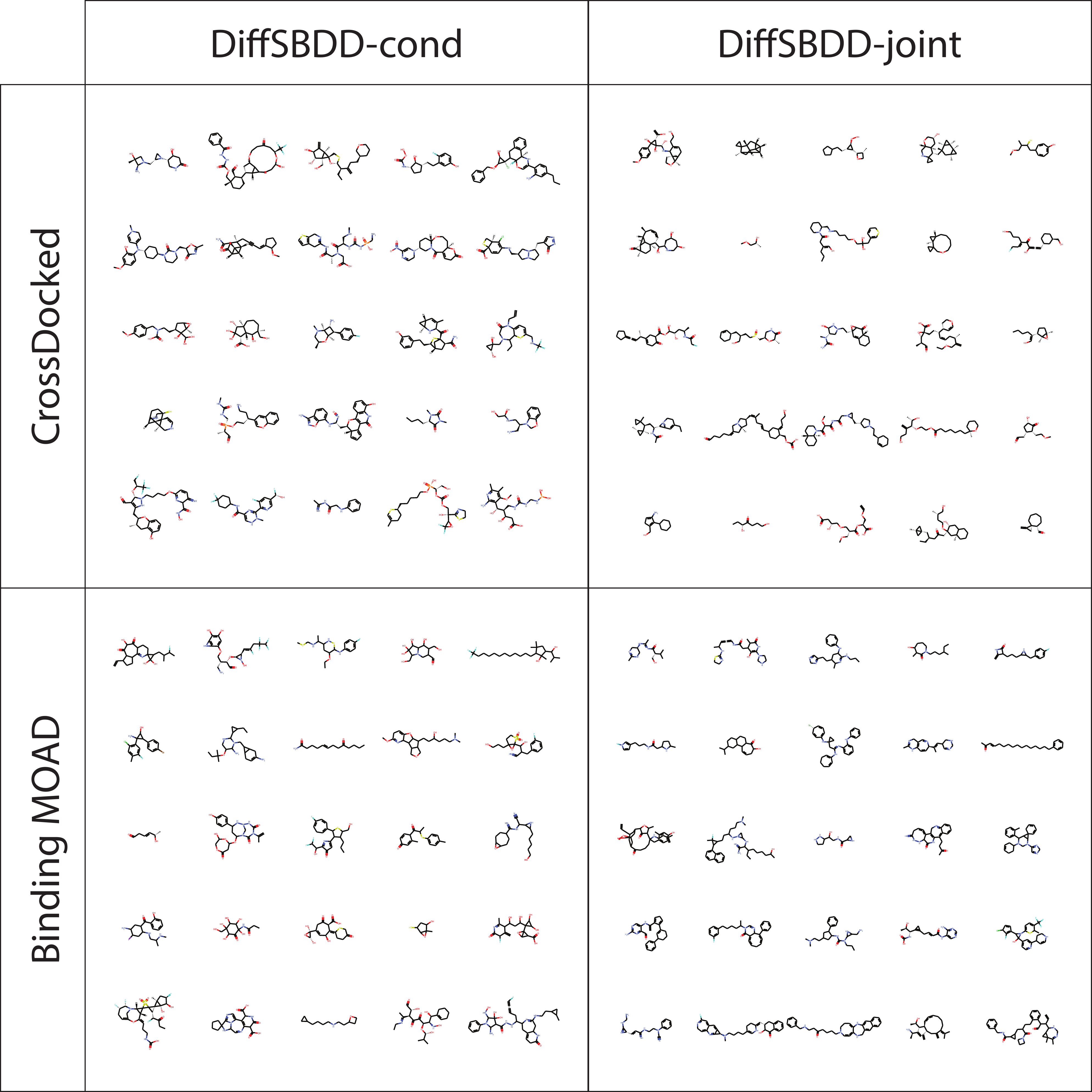}
     \caption{Randomly selected samples of generated molecules.}
     \label{fig:random_mols}
\end{figure}

Randomly selected molecules generated with different models are presented in Figure \ref{fig:random_mols}.

\section{Inpainting implementation}
\label{supp:inpainting_design}

\begin{figure}[h]
    \centering
    \includegraphics[width=\textwidth]{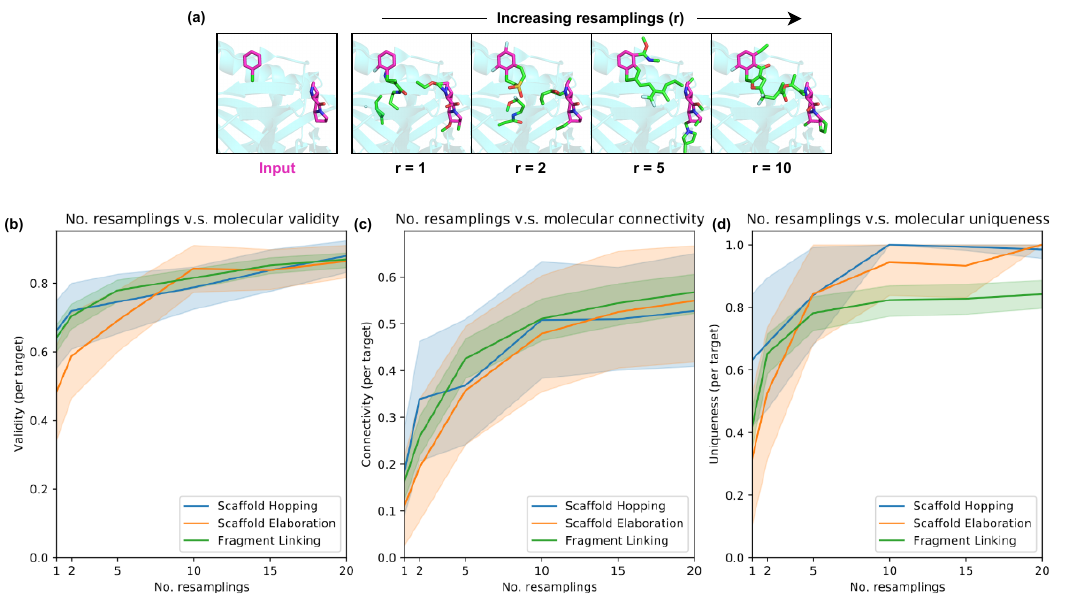}
    \caption{(a) Importance of high resamplings. Effect of the number of resamplings on molecular validity (b), connectivity (c) and uniqueness (d).}
    \label{fig:resampling_metrics}
\end{figure}

\subsection{Implementation of fragment merging experiments}
\label{app:fragment_merging}

Fragment merging is the task of combining fragments with an overlapping binding site~\cite{li2020applicationofFBDD}. For this example, instead of masking existing molecules, we instead take fragments from an experimental fragment screen against the non structural protein 3 (NSP3) from SARS-CoV-2~\cite{schuller2021fragment_screen}. Using two fragments as input (PDB entries 5rue and 5rsw) we successfully replicate the fragment merge performed in \citet{gahbauer2023iterative} which was accomplished using the chemoinformatics-based approach \textit{Fragmenstein}\footnote{https://github.com/matteoferla/Fragmenstein}. To accomplish this, instead of masking out and reinserting atoms, we instead choose to fix all atoms during generation except the atom on each fragment cloest to the other. We need perform $t=200$ steps of the DiffSBDD-\textit{diversify} procedure to allow the model to arrange the atom positions as well as change the atom types. All PDB files were already structurally aligned.

\subsection{Quantitative evaluation of inpainting for the whole Binding MOAD test set}\label{supp:quantitative_inpainting}

For all experiments across the whole test set, we perform automatic masking of atoms which are to be fixed. For scaffold elaboration, we extract the Bemis-Murcko scaffold \cite{bemis1996scaffolds} using RDKit and compute a binary mask to fix the scaffold, while functional groups are redesigned. For scaffold hopping, we simply take the inverse of the mask used for scaffold elaboration. For linker design, we fragment each molecule in the test set in multiple ways as in \citet{imrie2020delinker, igashov2022equivariant}.
To benchmark against DiffLinker, we use the model weights and protocol as described in \citet{igashov2022equivariant} except we give the ground-truth linker size as input, rather than predict it using the auxillary model, for fairness.
In small-scale experiments where finer control is desirable (e.g. as in the fragment merging example described above), the binary mask can be defined manually.

\section{Optimization}
\label{supp:optimisation}

\begin{figure}[h!]
    \centering
    \includegraphics[width=\textwidth]{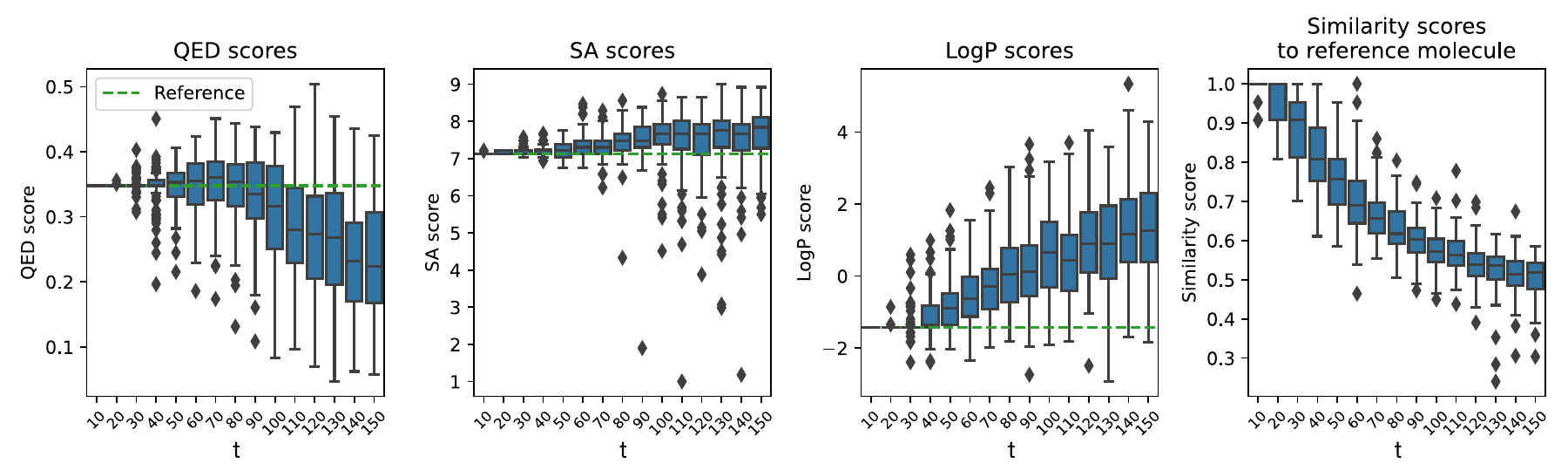}
    \caption{Effect of number of noising/denoising steps on molecule properties.}
    \label{fig:noise_denoise}
\end{figure}

We demonstrate the effect the number of noising/denoising steps ($t$) has on various molecular properties in Figure \ref{fig:noise_denoise}. We test all values of $t$ at intervals of 10 steps and 200 molecules are sampled at every timestep. Note this does not allow for explicit optimization of any particular property unless combined with the evolutionary algorithm, as shown in Figure \ref{fig:methods}D (all plots are for PDB entry 5NDU \cite{barone2020designed}).

\section{Related Work}

\paragraph{Diffusion Models for Molecules}
Inspired by non-equilibrium thermodynamics, diffusion models have been proposed to learn data distributions by modeling a denoising (reverse diffusion) process and have achieved remarkable success in a variety of tasks such as image, audio synthesis and point cloud generation~\cite{kingma2021variational,kong2021diffwave,luo2021diffusion}. Recently, efforts have been made to utilize diffusion models for molecule design~\cite{du2022molgensurvey}. Specifically, \citet{hoogeboom2022equivariant} propose a diffusion model with an equivariant network that operates both on continuous atomic coordinates and categorical atom types to generate new molecules in 3D space. Torsional Diffusion~\cite{jing2022torsional} focuses on a conditional setting where molecular conformations (atomic coordinates) are generated from molecular graphs (atom types and bonds).
Similarly, 3D diffusion models have been applied to generative design of larger biomolecular structures, such as antibodies~\cite{luo2022antigen} and other proteins~\cite{anand2022protein,trippe2022diffusion}.

\paragraph{Structure-based Drug Design}
Structure-based Drug Design (SBDD) \cite{ferreira2015SBDD2, anderson2003SBDD3} relies on the knowledge of the 3D structure of the biological target obtained either through experimental methods or high-confidence predictions using homology modelling \cite{kelley2015phyre2}. Candidate molecules are then designed to bind with high affinity and specificity to the target using interactive software \cite{kalyaanamoorthy2011software} and often human-based intuition \cite{ferreira2015SBDD2}. Recent advances in deep generative models have brought a new wave of research that model the conditional distribution of ligands given biological targets and thus enable \textit{de novo} structure-based drug design. Most of previous work consider this task as a sequential generation problem and design a variety of generative methods including autoregressive models, reinforcement learning, etc., to generate ligands inside protein pockets atom by atom~\cite{drotar2021structure,luo20213d,li2021structure,peng2022pocket2mol}. Most recent work explore the use of diffusion models in structure-based drug design~\citep{guan20233d, lin2022diffbp, guan2023decompdiff}.

\paragraph{Geometric Deep Learning for Drug Discovery}
Geometric deep learning refers to incorporating geometric priors in neural architecture design that respects symmetry and invariance, thus reduces sample complexity and eliminates the need for data augmentation~\cite{bronstein2021geometric}. It has been prevailing in a variety of drug discovery tasks from virtual screening to \textit{de novo} drug design as symmetry widely exists in the representation of drugs. One line of work introduces graph and geometry priors and designs message passing neural networks and equivariant neural networks that are permutation-, translation-, rotation-, and reflection-equivariant, respectively~\cite{duvenaud2015convolutional,gilmer2017neural,satorras2021n,lapchevskyi2020euclidean,du2022se}. These architectures have been widely used in representing biomolecules from small molecules to proteins~\cite{atz2021geometric} and solving downstream tasks such as molecular property prediction~\cite{schutt2018schnet,klicpera2020directional}, binding pose prediction~\cite{stark2022equibind}, transition state sampling~\cite{duan2023accurate}, and molecular dynamics~\cite{batzner20223,holdijk2022path}. Another line of work focuses on generative design of new molecules~\cite{du2022molgensurvey}. More specifically, molecule design is formulated as a graph or geometry generation problem, following either a one-shot generation strategy that generates graphs (atom and bond features) in one step or attempting to generate atoms and bonds sequentially.

\end{appendices}

\end{document}